\documentclass{aa}  

\usepackage{graphicx}
\usepackage{txfonts}
\begin{document} 

   \title{Dust properties and star formation of approximately a thousand local galaxies\thanks{Full Tables A.1, A.2, B.1, and B.2 are only available at the CDS via anonymous ftp to cdsarc.u-strasbg.fr (130.79.128.5) or via http://cdsweb.u-strasbg.fr/cgi-bin/qcat?J/A+A/.}}

   \author{S.~Lianou
         \inst{1,2}
         \and          
            P.~Barmby 
             \inst{3} 
           \and
           A.~A.~Mosenkov
             \inst{4,5}
           \and
           M.~Lehnert
             \inst{6}
           \and
           O.~Karczewski
             \inst{1}
          }

   \institute{Departement d'Astrophysique/IRFU, CEA, Universit\'{e} Paris-Saclay, Bat. 709, F-91191 Gif-sur-Yvette, France\\
     \email{sophia.thl@gmail.com}
    \and
     IAASARS, National Observatory of Athens, Penteli 15236, Greece
    \and
     Department of Physics \& Astronomy, University of Western Ontario, London, ON N6A 3K7, Canada
    \and
     Central Astronomical Observatory of RAS, Pulkovskoye Chaussee 65/1, 196140, St. Petersburg, Russia
    \and
    Sterrenkundig Observatorium, Department of Physics and Astronomy, Universiteit Gent Krijgslaan 281 S9, B-9000 Gent, Belgium
    \and
    Sorbonne Universit\'{e}, CNRS UMR 7095, Institut d'Astrophysique de Paris, 98 bis bd Arago, 75014 Paris, France
   }

   \date{Received October 31, 2018; accepted June 05, 2019}

  \abstract
      {}
      {We derived the dust properties for 753 local galaxies and examine how these relate to some of their physical properties. We present the derived dust emission properties, including model spectral energy distribution (SEDs), star formation rates and stellar masses, as well as their relations.}
      {We modelled the global dust-SEDs for 753 galaxies, treated statistically as an ensemble within a hierarchical Bayesian dust-SED modelling approach, so as to derive their infrared (IR) emission properties. To create the observed dust-SEDs, we used a multi-wavelength set of observations, ranging from near-IR to far-IR-to-submillimeter wavelengths. The model-derived properties are the dust masses (M$_{dust}$), the average interstellar radiation field intensities (U$_{av}$), the mass fraction of very small dust grains (`QPAH' fraction), as well as their standard deviations. In addition, we used mid-IR observations to derive star formation rates (SFR) and stellar masses, quantities independent of the dust-SED modelling. }
      {We derive distribution functions of the properties for the galaxy ensemble and as a function of galaxy type. The mean value of M$_{dust}$ for the early-type galaxies (ETGs) is lower than that for the late-type and irregular galaxies (LTGs and Irs, respectively), despite ETGs and LTGs having stellar masses spanning across the whole range observed. The U$_{av}$ and `QPAH' fraction show no difference among different galaxy types. When fixing U$_{av}$ to the Galactic value, the derived `QPAH' fraction varies across the Galactic value (0.071). The specific SFR increases with galaxy type, while this is not the case for the dust-specific SFR (SFR/M$_{dust}$), showing an almost constant star formation efficiency per galaxy type. The galaxy sample is characterised by a tight relationship between the dust mass and the stellar mass for the LTGs and Irs, while ETGs scatter around this relation and tend towards smaller dust masses. While the relation indicates that M$_{dust}$ may fundamentally be linked to M$_{\star}$, metallicity and U$_{av}$ are the second parameter driving the scatter, which we investigate in a forthcoming work. We used the extended Kennicutt-Schmidt (KS) law to estimate the gas mass and the gas-to-dust mass ratio (GDR). The gas mass derived from the extended KS law is on average $\sim$20\% higher than that derived from the KS law, and a large standard deviation indicates the importance of the average star formation present to regulate star formation and gas supply. The average GDR for the LTGs and Irs is 370, and including the ETGs gives an average of 550. }
      {}

   
   \maketitle
%

\section{Introduction}

The stellar mass of a galaxy bears the imprints of galaxy evolution through hierarchical growth \citep{Oser2010, Somerville2015}. Moreover, it encodes the amount of gas locked in its long-lived stars. Together with the stellar mass (M$_{\star}$), the star formation rate (SFR) of galaxies is fundamental in understanding their star formation histories (SFHs) over cosmic times \citep{Madau2014}. Star formation (SF) plays a significant role in shaping the interstellar medium (ISM) in galaxies \citep{Hopkins2012,Ceverino2009,Lehnert1996,Heckman1993}. The relation of SF to the gas available in a galaxy is imprinted in the Kennicutt-Schmidt law \citep[KS law;][]{Kennicutt1998,SLAW}, while M$_{\star}$ also plays a crucial role in regulating SF both on local and global scales in galaxies \citep[extended Schmidt law;][]{Shi2011,Rahmani2016}.

Understanding the link between SF and the ISM has the aim of uncovering the evolutionary history of galaxies across different environments and redshift \citep{Madau2014,Somerville2012,Granato2000,Pei1999}. Coupling SF to the ISM in galaxies requires knowledge of their energy emitted across the electromagnetic spectrum. Integrated light analyses include many physical processes acting at different spatial scales, which are imprinted in the shape of the spectral energy distribution (SED) of the galaxy \citep{Conroy2013,Walcher2011}. Crucial to the shape of an SED within a galaxy is the presence of dust and the dust grain size distribution. In panchromatic SED analyses, the energy balance of the ultraviolet-to-optical starlight absorbed by the dust and re-emitted as thermal radiation in the infrared (IR) is assumed \citep[e.g.,][]{Noll09,dacunha08}. In dust-SED analyses, the IR-to-submillimeter part of the spectrum is used to infer the properties of the dust emission \citep[e.g.,][]{Dale2001,Siebenmorgen2007,Compiegne2011,Kelly2012}.

We made use of the wealth of multi-wavelength information for a statistically large sample of galaxies in the local Universe with the aim of gaining insights on the relation between the fundamental properties describing the galaxies and their evolution, linking their IR-to-submillimeter (submm) emission to their past-to-present average SFHs, as decoded in their SFR and M$_{\star}$. The relation of SFR and M$_{\star}$ reveals a tight sequence for the star-forming galaxies \citep{Brinchmann2004,Elbaz2007,Noeske2007}, with a slope consistent with $\sim$one independent of redshift, while its normalisation changes to reflect a higher SFR at higher redshift at a given stellar mass \citep{Elbaz2011}.

We drew a sub-sample of galaxies from the DustPedia project, which provides public access to a large photometric and imaging dataset for 875 galaxies in the local Universe \citep[][]{Clark2018Dust}. Our study combined the homogeneous treatment of the photometry with a dust-SED model that uses a hierarchical Bayesian framework, allowing us to model the largest sample of local galaxies yet modelled as an ensemble and to place statistical constraints on the derived dust emission properties. The galaxy sample and its properties are presented in Section 2, while the dust-SED model in Section 3. We present the resulting modelled dust-SEDs in Section 4. The model-derived properties, that is, average interstellar radiation field (ISRF) intensities (U$_{av}$), dust masses (M$_{dust}$), the mass fraction of very small grains (QPAH), as well as their correlation, are shown in Section 5. The relation between SF and model-derived properties are presented in Section 6, including the KS law and the extended KS law. The model-derived and calibration-derived physical parameters, as well as the model dust-SEDs, are made publicly available here. Tables A.1 and A.2 in the Appendix A serve as examples to their contents. In addition to the modelling described in Section 3, the model output regarding the modified black body fits to the data are given in the Appendix B. We summarise our findings in Section 7. 

\section{Analysis}

\subsection{Galaxy sample and photometric measurements}

The details of the DustPedia photometric data that we used for this study are described in full in \citet{Clark2018Dust}, and a brief description of the photometry follows. The galaxy sample consists of 753 galaxies, which is a sub-sample drawn from the DustPedia photometric catalogue \citep{Clark2018Dust}. The initial DustPedia photometric catalogue consists of 875 galaxies with aperture-matched photometric measurements spanning a large range of bands, with a maximum of 42 wavelengths ranging from the ultraviolet (UV) to the submm, and an average of 25 bands.  

At the initial stage of the photometry, we removed any contamination from foreground stars in UV to mid-IR bands, and then we identified and removed any large-scale background structures (such as cirrus, airglow, etc) by means of 2D polynomial fit. Subsequently, for each given target galaxy, apertures of elliptical shape were used to fit the source in each band; the apertures for every band were then combined to yield a `master' elliptical aperture for that target (it should be noted that aperture dimensions, as recorded and applied in each band, are adjusted to account for the bands' point spread function, PSF). The source flux is then measured in each band using this master elliptical aperture (including consideration of partial pixels). Local background subtraction was performed by taking the iteratively sigma-clipped mean of the pixel values contained within an elliptical sky annulus (with the same axial ratio and position angle as the source aperture, extending from 1.25 to 1.5 times the source aperture semi-major axis). Aperture corrections were applied to correct for the fraction of the source flux that fell outside the aperture due to the effect of each band's PSF. At wavelengths shorter than 10$\mu$m, we corrected the fluxes for foreground Galactic extinction using the IRSA Galactic Dust Reddening and Extinction Service, which uses the prescription of \citet{Schlafly2011}.

Photometric uncertainties were estimated by placing random sky apertures across the map surrounding the source aperture. The variation between the flux measured in these random sky apertures (determined by taking the iteratively sigma-clipped standard deviation of the flux measured in each) encompasses instrumental noise, confusion noise, and sky noise. This uncertainty was then added in quadrature to the calibration uncertainty for the instrument in question to yield the final photometric uncertainty
\footnote{The calibration uncertainty assumed in \citet{Clark2018Dust} is different than that assumed internally in the hierarchical Bayesian dust-SED model \citep[see Table 1 in][for the calibration uncertainties assumed in the dust-SED model]{Lianou2018a}. Therefore, we have made the necessary corrections to all bands, so as to use the calibration uncertainties assumed here in the dust-SED model, and not those assumed in \citet{Clark2018Dust}.}.

The aperture-matched photometry was combined with existing legacy photometry from the InfraRed Astronomical Satellite \citep[IRAS;][]{Neugebauer1984} Scan Processing and Integration tool (SCANPI), and Planck \citep{Planck2011} 2nd Catalogue of Compact Sources \citep[Planck CCS2;][]{Planck2016}. This supplementary photometry was flagged in accordance with the flagging procedures for the DustPedia photometry. The photometric measurements reported in \citet{Clark2018Dust} are used to model the dust-SED of the 753 galaxies. We use the D25 major axis diameter, at which the optical (in the B band) surface brightness falls beneath 25~mag\,arcsec$^{-2}$ \citep[adopted from][]{Clark2018Dust}, in order to derive the area, with which we normalise to derive surface densities of physical quantities, such as the surface density of SFR. The adopted D25 values for all galaxies are listed in Table A.1 in the Appendix A.

To select this galaxy sub-sample for the dust-SED modelling, we make use of the flags reported in the initial DustPedia photometric catalogue \citep['global' flags and/or individual-band flags;][]{Clark2018Dust}. If a galaxy's photometry is reported with a 'global' flag, then this means that there is a contamination by another source affecting a large number of wavelengths. There is a total of 83 galaxies reported with such a 'global' flag in \citet{Clark2018Dust}, and these galaxies have been excluded from the SED modelling. An additional 39 galaxies that do not have any far-IR constraint and with an SED dominated by strong mid-IR emission has been excluded from the SED modelling of the ensemble. After this selection process, there are 753 galaxies left (i.e., 89\% of the original sample) and modelled as an ensemble on global scales, using the dust-SED model described in Section 3.

\begin{table}
  \tiny
      \begin{minipage}[t]{\columnwidth}
      \caption[]{Bands used for the SED modelling. }
      \label{sl_table1} 
      \renewcommand{\footnoterule}{}
      \begin{tabular}{lllc}
\hline\hline
Telescope/Filter\footnote{Filter here is defined to mean either the instrument used, or the camera used, or the filter used, depending on the telescope facility.}       
&$\lambda$      &FWHM\footnote{{\bf References for FWHM.--} these are the same as in \citet{Lianou14}, while for Planck\,HFI this is given in \citet{Planck2014VII}.}
&Number of galaxies                                                                \\
                            &($\mu$m)               &(arcsec)               &       \\         
\hline                                  
2MASS/J                &1.25                   &2.5                     &732       \\
2MASS/H                &1.65                   &2.5                     &708       \\
2MASS/K$_{S}$           &2.17                   &2.5                    &729        \\
{\it WISE}/W1          &3.4                    &8.4                    &715       \\
{\it WISE}/W2          &4.6                    &9.2                    &715       \\
{\it Spitzer}/IRAC     &8.0                    &2.0                    & 99       \\
{\it WISE}/W3          &12                     &11.4                   &735       \\
{\it WISE}/W4          &22                     &18.6                   &740       \\
{\it Herschel}/PACS    &70                     &5.8                    &125         \\
{\it Herschel}/PACS    &100                    &7.1                    &501         \\
{\it Herschel}/PACS    &160                    &11.2                   &522         \\
{\it Herschel}/SPIRE   &250                    &18.2                   &698         \\
{\it Planck}/HFI       &350                    &260.0                  &360        \\
{\it Herschel}/SPIRE   &350                    &25.0                   &698         \\
{\it Herschel}/SPIRE   &500                    &36.4                   &692         \\
{\it Planck}/HFI       &550                    &281.0                  &251        \\
{\it Planck}/HFI       &850                    &290.0                  &181        \\
\hline
\end{tabular} 
\end{minipage}
\end{table}
%
The dust-SED model requires the use of observations in the near-IR to submm bands, listed in Table~\ref{sl_table1}, using several facilities: 2MASS \citep{Skrutskie06}, {\it WISE} \citep[][]{Wright10}, {\it Spitzer} IRAC \citep{Fazio04},  {\it Herschel} \citep{Pilbratt2010} PACS \citep{Poglitsch10} SPIRE \citep{Griffin10}, and {\it Planck}/HFI \citep{Lamarre2010}. The number of galaxies detected per band is listed in Table~\ref{sl_table1}. For each galaxy, if its individual band photometric measurements was accompanied with a minor/major flag, this band has been excluded from the SED fitting of that galaxy (hence the number of galaxies per band is less than 753). For the 753 galaxies considered here, the median number of bands used in the SED modelling is twelve. The minimum number of bands used is four (in only one galaxy, NGC\,4636) and the maximum number of bands used is seventeen, that is, the maximum possible (for four galaxies: NGC\,3256, NGC\,3982, NGC\,6946, UGC\,12160), while another three galaxies have only six bands (PGC\,029653; NGC\,2974; ESO\,411-013). There are 740 galaxies with {\it WISE}\,22$\mu$m observations, and 99 galaxies with {\it Spitzer} IRAC\,8$\mu$m observations. We have included the {\it Spitzer} IRAC\,8$\mu$m due to the unique constraint to the mid-IR part of the dust-SED it provides \citep{Draine07}.

%
 \begin{table}
      \begin{minipage}[t]{\columnwidth}
      \caption[]{Galaxy number per Hubble type considered in the dust-SEDs.}
      \label{sl_table2} 
      \renewcommand{\footnoterule}{}
      \begin{tabular}{ccccc}
\hline\hline
Type bin                   &All            &ETGs               &LTGs                  &Irrs            \\ 
\hline
-5$\leq$T$\leq$10          &753            &235                &340                   &178             \\
\hline
-5$\leq$T<-4               &...            &53                 &0                     &0               \\
-4$\leq$T<-3               &...            &11                 &0                     &0               \\
-3$\leq$T<-2               &...            &53                 &0                     &0               \\
-2$\leq$T<-1               &...            &78                 &0                     &0               \\
-1$\leq$T$\leq$0           &...            &40                 &0                     &0               \\
\hline
0<T<1                      &...            &0                  &42                     &0               \\
1$\leq$T<2                 &...            &0                  &36                     &0               \\
2$\leq$T<3                 &...            &0                  &41                     &0               \\
3$\leq$T<4                 &...            &0                  &86                     &0               \\
4$\leq$T<5                 &...            &0                  &42                     &0               \\
5$\leq$T$\leq$6            &...            &0                  &93                     &0               \\
\hline
6<T$\leq$7                 &...            &0                  &0                     &53              \\
7<T$\leq$8                 &...            &0                  &0                     &37              \\
8<T$\leq$9                 &...            &0                  &0                     &31              \\
9<T$\leq$10                &...            &0                  &0                     &57              \\
\hline
\end{tabular} 
\end{minipage}
\end{table}
 %
The adopted galaxy sample is characterised by a large variety of galaxy types: 235 early-type galaxies (ETGs), including elliptical and lenticular galaxies; 340 late-type spiral galaxies (LTGs); 178 irregular galaxies (Irs), including Magellanic irregulars. By selection, the DustPedia galaxy sample contains nearby galaxies (within $\sim$40Mpc) with angular sizes (optical diameters) larger than 1$\arcmin$ and observed with the {\it Herschel} Space Observatory \citep{Clark2018Dust}. In what follows, in order to characterise the galaxy type, we use the revised Hubble type (T), a numerical scheme which assigns numbers to Hubble types, spanning the range from -5 (elliptical galaxies) to 10 (irregular galaxies). The numerical values T of the Hubble stage is retrieved from HyperLeda \footnote{\url{http://leda.univ-lyon1.fr/}} \citep{Makarov2014} as described in \citet{Clark2018Dust}. For our subsequent analysis, we define three broad galaxy type bins, assuming that: ETGs have T$\le$0 and assign in all figures a red colour to galaxies of this type; LTGs have 0$<$T$\le$6 and assign in all figures a blue colour to galaxies of this type; Irs have T$>$6 and assign a lighter blue colour to galaxies of this type. The ETGs comprise 32\% of the total galaxy sample, the LTGs comprise 45\% of the total galaxy sample, and the Irs comprise 24\% of the total galaxy sample. Table~\ref{sl_table2} lists the number of galaxies per type considered for the construction of the median model dust-SEDs. The derived model dust-SEDs of each galaxy in the ensemble are given in Table A.2 in Appendix A.

\subsection{Derivation of M$_{\star}$ and SFR}

Deriving M$_{\star}$ for a galaxy relies on several methodologies each subject to uncertainties, for example, assumptions on SFH, the initial mass function, the treatment of the asymptotic giant branch phase of stars \citep[][and references therein]{Courteau2014}. Galaxies contain stellar populations dominated by low- and intermediate-mass stars (main sequence or evolved stars), forming the bulk of the mass of a galaxy \citep{Chabrier2003}, and most of their light is emitted in the near-IR ($\sim$1-5$\mu$m). Given that stellar mass determinations from near-to-mid-IR bands are less affected by stellar population variations, we opt to use {\it WISE} 3.4$\mu$m observations to derive M$_{\star}$ on global scales for the 753 galaxies studied here. We adopt the calibration in \citet[][their eq.~2]{Wen2013}, which relates M$_{\star}$ to the {\it WISE} 3.4$\mu$m for a sample of $\sim$5$\times$10$^{5}$ galaxies drawn from the MPA-JHU Sloan Digital Sky Survey catalogue.

Deriving SFRs from observations is also non-trivial, and several indicators exist and probe different timescales for star formation \citep{Calzetti13,Kennicutt12,Calzetti2007,Salim2007}. In order to derive the global SFR for the 753 galaxies, we use the {\it WISE} 22$\mu$m band. We adopt the calibration from \citet[][their equation 6]{Cluver2017}, who demonstrate that the {\it WISE} 22$\mu$m and 12$\mu$m bands provide SFRs in excellent agreement with measurements combining obscured and unobscured SFR tracers. While the WISE 12$\mu$m SFR shows relatively less scatter (0.15dex) than the one based on {\it WISE} 22$\mu$m (0.18dex), the former may suffer from silicate absorption features. Using the {\it WISE} 22$\mu$m band to derive SFRs for ETGs means that this SFR tracer probes a timescale difference as compared to the SFR for LTGs or Irs. This is because stellar populations in ETGs are dominated by low-mass main sequence and evolved stars, while their SFR levels are on average low and rather localised. Ongoing or recent SF in ETGs, at timescales less than $\sim$100-300\,Myr, have been reported in some cases \citep{Moellenhoff81,Ford2013}, but evolved stars are the main contributors to the emission in the mid-IR bands \citep{Jarrett2013,Petty2013}.

%
\begin{table}
  \tiny
      \begin{minipage}[t]{\columnwidth}
      \caption[]{Median values of the calibration-derived SFR and M$_{\star}$, for the ensemble of galaxies and per galaxy type. The quoted values in the parenthese denote the median absolute deviation.}
      \label{sl_table3} 
      \renewcommand{\footnoterule}{}
      \begin{tabular}{ccccc}
\hline\hline
Property                              &All          &ETGs               &LTGs                  &Irrs            \\ 
\hline\hline
SFR (M$_{\sun}$/yr)                   &0.30(0.28)    &0.11(0.09)      &0.81(0.69)           &0.17(0.14)          \\
M$_{\star}$(10$^{10}$M$_{\sun}$)         &0.75(0.69)   &1.07(0.97)      &1.63(1.35)           &0.15(0.13)           \\
\hline
\end{tabular} 
\end{minipage}
\end{table}
%
To correct for the contribution to the mid-IR band from evolved stars, and prior to applying the calibration of \citet[][]{Cluver2017}, we use the method outlined in \citet{Temi2009} and \citet{Davis2014}. More specifically, \citet{Davis2014} obtain for passive galaxies, dominated by low-mass evolved stars, a relation between their K$_{S}$ and {\it WISE} 22$\mu$m luminosities, in order to characterise the contribution of these stellar populations to the {\it WISE} 22$\mu$m emission \citep[eq.~1,][]{Davis2014}. Then, they subtract the contribution of these stellar populations to the {\it WISE} 22$\mu$m emission, in order to derive the emission related to SF \citep[eq.~2,][]{Davis2014}. Here, we use the 2MASS/K$_{S}$ band along with the relation in \citet[][their eq.~1]{Davis2014} to estimate the emission in the mid-IR from the evolved stars. We then subtract this from the {\it WISE} 22$\mu$m emission \citep[using eq.~2 in ][]{Davis2014}. The emission thus remained in {\it WISE} 22$\mu$m is associated with SF, and this we use with the relation of \citet{Cluver2017} to derive the SFRs. We apply this correction to all galaxies in our sample, as all galaxies contain evolved stars and Table A.1 in the Appendix A lists the derived SFRs and M$_{\star}$. Table~\ref{sl_table3} lists the median (and median absolute deviation) value for the SFR and M$_{\star}$ of the galaxy ensemble. The LTGs show a higher SFR as compared to the ETGs and Irs. The stellar mass of the ETGs and LTGs are comparable, while the Irs lie on the lower mass regime. 

\section{Modelling the spectral energy distributions}  

The observed dust-SEDs for the 753 galaxies are modelled with a dust-SED model tool \citep{Galliano2018} that applies an hierarchical Bayesian (HB) approach and incorporates the properties of the \citet[][hereafter THEMIS]{Jones2017} dust grain model. A simplified version of the dust-SED model was used in \citet[][hereafter G11]{Galliano2011}, wherein the HB approach or the THEMIS dust grain model were not included. A comparison between the dust-derived properties adopting either the THEMIS dust grain composition or the amorphous carbon dust grain composition of G11 (i.e., their `AC' model) has been performed in \citet[][]{Lianou2018a} and the result of this comparison shows that the derived properties have similar values within the uncertainties. In the current study we adopt the THEMIS dust grain composition. 

Dust grains are characterised by their chemical composition, size distribution, shape, and abundances \citep{Savage1979,Zubko2004,Draine2009}. The main principle of the THEMIS model is that dust evolves in response to the physical conditions of its local environment and is not characterised by the same properties everywhere; therefore, dust has different chemical composition, structure, shape, according to the physical conditions exposed, which in turn affects its optical properties \citep{Jones2017}. The THEMIS model is built, as much as possible, upon laboratory measurements of dust material analogues to the interstellar dust, and modifications necessary to provide better fits to the observed interstellar dust properties have been made \citep[][]{Jones2013}. The dust grain composition in the THEMIS model is a mixture of amorphous silicates, with iron and iron-sulphide nano-inclusions, and hydrogenated amorphous carbon materials (a-C(:H)), while the size distribution of the different dust populations are shown in Fig.~2 of \citet[][see also their Section 5]{Jones2013}. 

The model fits to the observed dust-SEDs are performed within a HB framework and the sample of 753 galaxies is modelled as an ensemble. This allows us to include in the ensemble galaxies with photometric measurements that have small signal-to-noise ratio and to handle several sources of uncertainties in a statistical way \citep[][]{Kelly2007,Kelly2012}. The physical components in the dust-SED model and the HB formalism are described in full in \citet{Galliano2018}. Briefly, the model's hierarchical structure means that there are multiple levels in the formulation of the prior distribution of the model parameters, which are combined to formulate a multivariate Student's-t distribution that depends on hyperparameters \citep[average of each of the parameters, $\mu$, and their covariance matrix, $\Sigma$, see eq.~19 in][]{Galliano2018}. The inference from the data occurs via Markov Chain Monte Carlo (MCMC) simulations, after which run the distribution of the hyperparameters becomes updated \citep[][]{Galliano2018}. The difference between Bayesian modelling and HB modelling is that in the latter case the uncertainties of the observational dataset (measurement or calibration errors, but also upper limits) is properly accounted for at all levels of the modelling \citep{Kelly2012,Kelly2007}, appropriate for handling statistical ensembles. The HB approach is important in the case of low signal-to-noise ratio sources within the ensemble, for which the prior distribution will have a much larger effect \citep{Kelly2012}. Even though there are several physical component options provided within the dust-SED tool to model the observed dust-SED, including fitting with modified black bodies\footnote{See Appendix B for results related to the modified black body fits to our galaxy sample.}, we opted for the ones we describe in what follows.

The dust-SED model we use here is the linear combination of two physical components: (i) a non-uniformly illuminated dust mixture (that of THEMIS), and (ii) of a stellar continuum modelled as a black body with temperature T=50000K \citep[see Sections 2.2.5 \& 2.2.6, respectively, in][]{Galliano2018}. The model is described with seven parameters: six for the non-uniformly illuminated dust component, and one for the stellar continuum. Hence, the seven model parameters are:
\begin{enumerate}
\item the dust mass, M$_{dust}$ (no limit on its range, see eq.~1 below for definition),
\item the minimum radiation field intensity, U$_{min}$, ranging between $\sim$-4.6 to $\sim$6.9 (see eq.~1 below),
\item the interval length of the radiation field intensity, $\Delta$U, ranging between 0 and $\sim$13.8 (see eq.~1 below),
\item the power law index, $\alpha$, describing the distribution of the dust mass per unit heating intensity and ranging between 1.0 and 2.5 (see eq.~1 below), 
\item  the `QPAH' fraction \citep[or q$_{i}^{PAH}$ in the notation of ][]{Galliano2018}, with values ranging from 0 to 0.9. In the THEMIS framework, the `QPAH' fraction refers to the mass fraction of hydrogenated amorphous carbon dust grains with sizes between 0.7nm to 1.5nm, where the Galactic value is 0.071,
\item  the fraction of charged 'PAHs', f$^{+}$, ranging from 0 to 1, 
\item  the bolometric luminosity of the stellar continuum component, L$_{\star}$ (no limit on its range).
\end{enumerate}
The average ISRF intensity, U$_{av}$, as defined in eq.~9 in \citet{Galliano2018}, is used to discuss the heating intensity. This quantity is used instead of the free parameters U$_{min}$, $\Delta$U and $\alpha$, as the latter may occasionally be subject to degeneracies \citep[][]{Galliano2018}. The average ISRF (or heating) intensity indicates the peak of the far--IR part of the SED, which is linked to the dust temperature \citep{Draine2009}. One assumption in the model is the abundance of the small grains (f$_{SG}$, with sizes less than 10nm, including carbon and silicate grains), which is kept fixed at a value three times lower than the Galactic one (i.e., the value used here is f$_{SG}$$\sim$0.026). The reason for this assumption is the quality of the fits in the mid-IR. We explored the quality of the fits using an abundance of the small grains similar to the Galactic value see Fig.~\ref{sl_figA1} in the Appendix A).

We derive dust masses using the phenomenological model described in \citet{Galliano2018}, as per \citet[][]{Dale2001} and \citet{Desert1990}. A mass element of the ISM is assumed to be illuminated by a non-uniform interstellar radiation field. The latter is described by a heating intensity {\it U}, with {\it U}\,=\,1 corresponding to the intensity of the solar neighbourhood, {\it U$_\sun$}\,=\,2.2 $\times$ 10$^{-5}$\,W\,m$^{-2}$. Then, the distribution of the dust masses per unit heating intensity is described by the power law over heating intensities:
\begin{equation}
\frac{dM_{dust}}{dU} \propto U^{- \alpha}, \,  with \, U_{min} < U < U_{min} + \Delta U .
\end{equation}
Integrating the above expression between U$_{min}$ to U$_{min} + \Delta U$ results in the total dust mass, M$_{dust}$, for the assumed mass element.

The model-output properties considered are the mean values of the probability density function{\bf. These were} derived from sampling 1.5$\times$10$^{5}$ random draws of MCMC simulations of the model fitted to the data.

\section{Model dust-SEDs}

 \begin{figure*}[th!]
  \centering
      \includegraphics[width=7.9cm,clip]{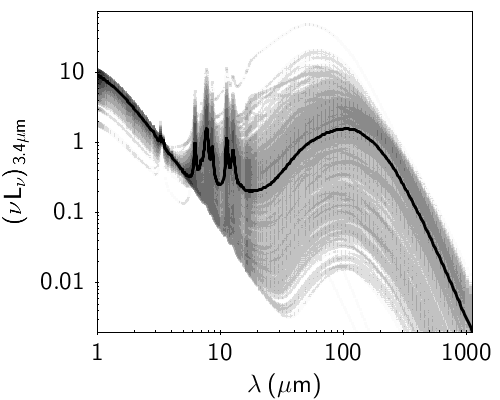}
      \includegraphics[width=7.9cm,clip]{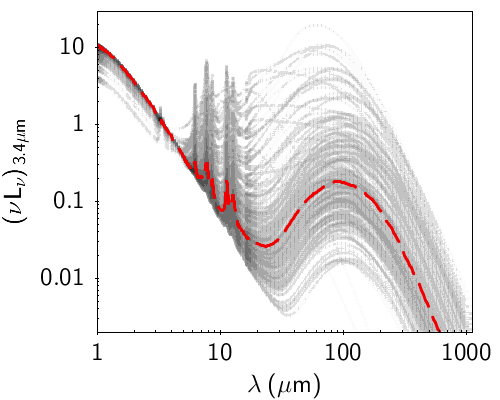}
      \includegraphics[width=7.9cm,clip]{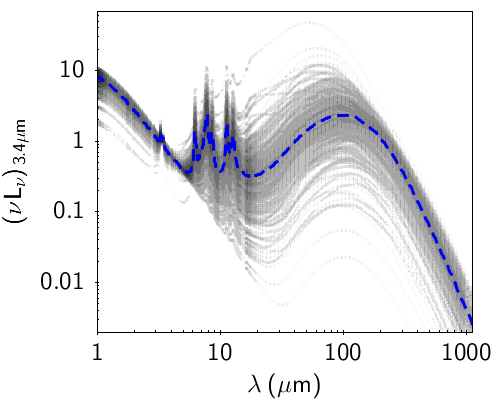}
      \includegraphics[width=7.9cm,clip]{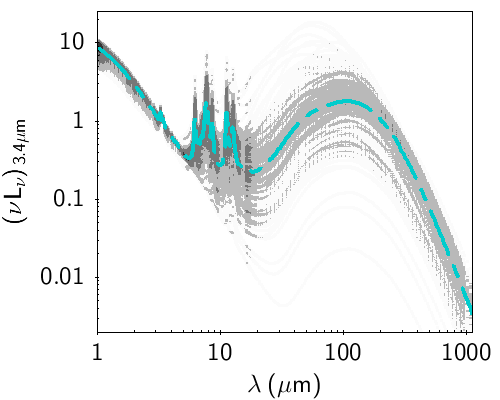}
      \caption{{\it Upper left panel}: Median model dust-SEDs for individual galaxies (grey lines). The black line represents the median model dust-SED of the whole (753 galaxies) sample. All dust-SEDs are normalised to the 3.4$\mu$m luminosity. 
        {\it Upper right panel}: Same as upper left panel, but for individual ETGs.
        {\it Left lower panel}: Same as upper left panel, but for individual LTGs.
        {\it Right lower panel}: Same as upper left panel, but for individual Irs. }
      \label{sl_figure1b}%
 \end{figure*}
%
 The model dust-SEDs for each galaxy are shown in Fig.~\ref{sl_figure1b}. The upper left panel shows the model dust-SEDs of each of the 753 galaxies separately (grey lines), and the model dust-SEDs of only the ETGs (upper right panel), the LTGs (lower left panel), and the Irs (lower right panel). Also shown for comparison is the median model dust-SED of the whole sample, or per galaxy type (solid lines). A zoom-in of the mid-IR part of the spectrum is shown in the Appendix A (Fig.~\ref{sl_figA2}). Each galaxy\footnote{Fig.~\ref{sl_figA3} in the Appendix A shows the model dust-SED density for selected individual galaxies.} is characterised by a variety of physical processes averaged on global galaxy scales, and the shapes of the model dust-SEDs show different mid-IR emission, and slightly hotter dust-SED for the ETGs. 

 \begin{figure}
  \centering
      \includegraphics[width=7.7cm,clip]{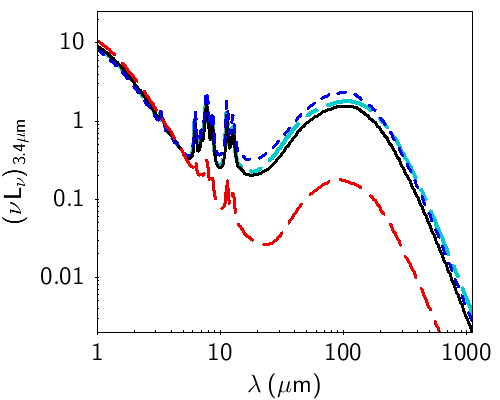}
      \includegraphics[width=7.5cm,clip]{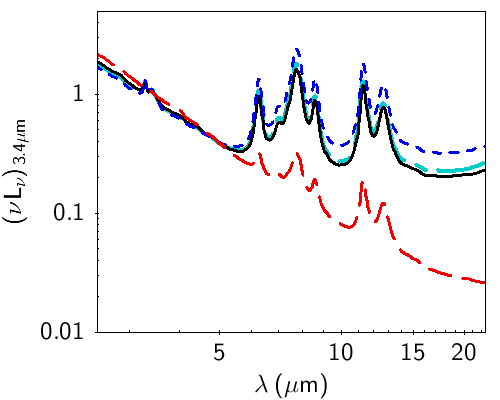}
      \caption{{\it Upper panel}: Median model dust-SED for the 753 galaxies (black solid line). The median model dust-SED separately for the ETGs, LTGs, and Irs is also shown with the dashed red, dashed blue, and dashed light-blue lines. All model dust-SEDs are normalised to the 3.4$\mu$m luminosity.
      {\it Lower panel}: Exact same as in the upper panel, but zooming into the mid-IR part of the spectrum, from 2.5$\mu$m to 22.5$\mu$m.}
      \label{sl_figure1}%
 \end{figure}
%
 The latter can be clearly seen in Fig.~\ref{sl_figure1}, where only the median model dust-SEDs are shown to facilitate easy comparison. The upper panel in Fig.~\ref{sl_figure1} shows the median model dust-SED for the whole galaxy sample (black solid lines). The median model dust-SEDs for the ETGs, LTGs, and Irs are also shown with the dashed red, blue, and light-blue lines, respectively. The lower panel in Fig.~\ref{sl_figure1} is a zoom-in of these median model dust-SEDs into the mid-IR part of the spectrum, from 2.5$\mu$m to 22.5$\mu$m. The median model dust-SEDs representing the LTGs and the Irs show a similar shape when compared to each other. For the Irs, the median model dust-SED of the 753 galaxies provides a very good representation of their median model dust-SED, when considering the full wavelength range or the mid-IR part only. The median model dust-SED of the ETGs tends to lower luminosities, as compared to the other two type of galaxies or the median over the ensemble.

\section{Results on model-derived properties}

\subsection{Distribution functions}

 \begin{figure*}
  \centering
      \includegraphics[width=6cm,clip]{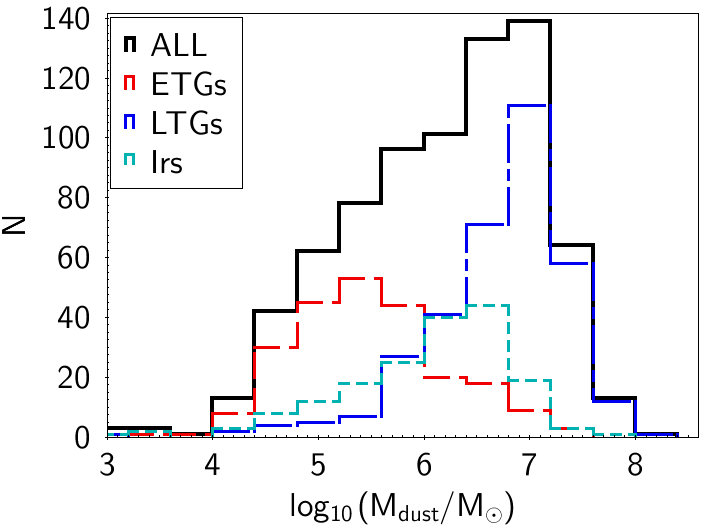}
      \includegraphics[width=6cm,clip]{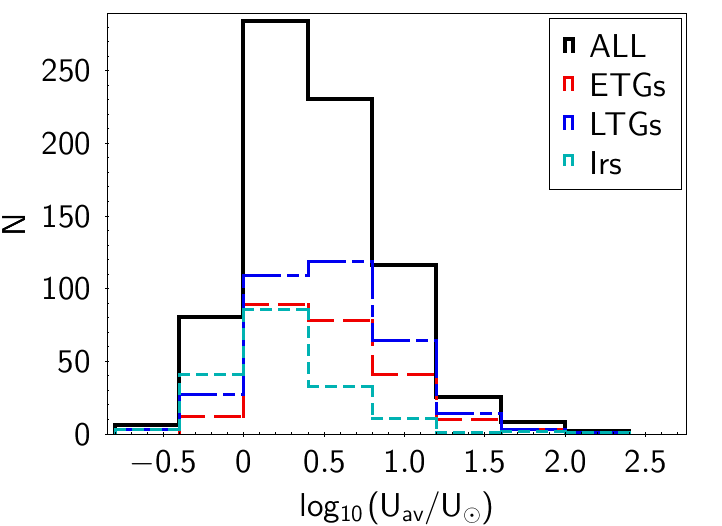}
      \includegraphics[width=6cm,clip]{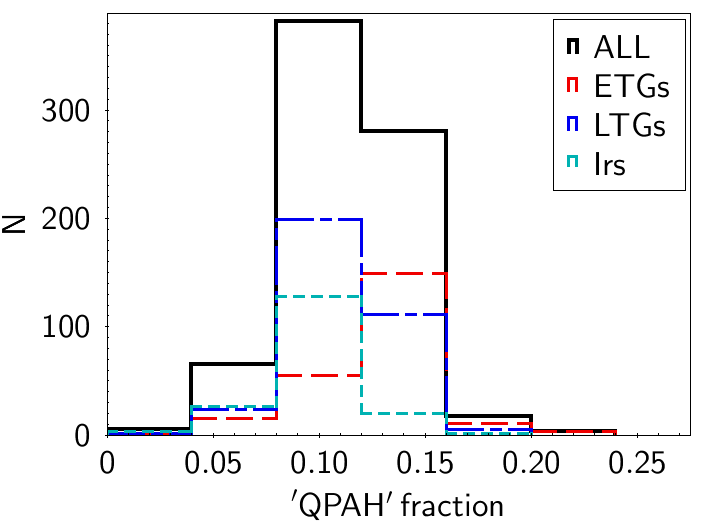}      
      \includegraphics[width=6cm,clip]{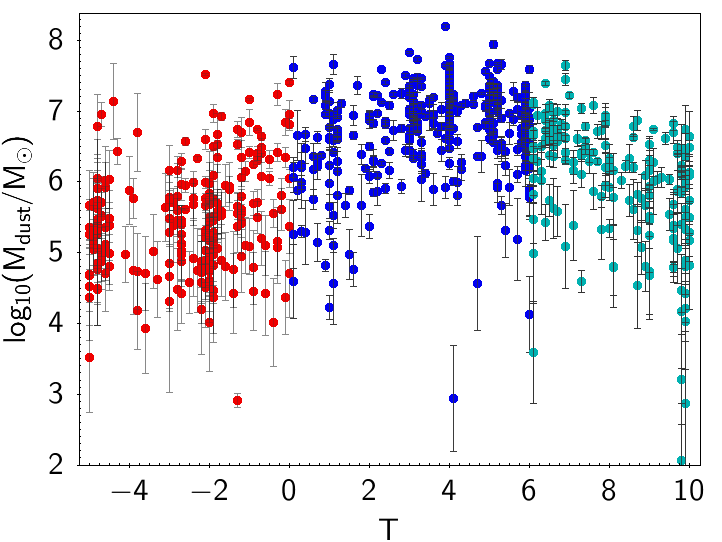}
      \includegraphics[width=6cm,clip]{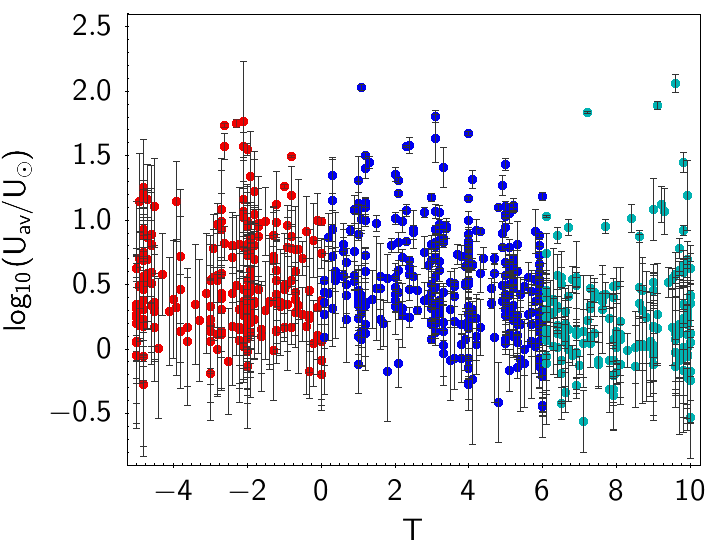}
      \includegraphics[width=6cm,clip]{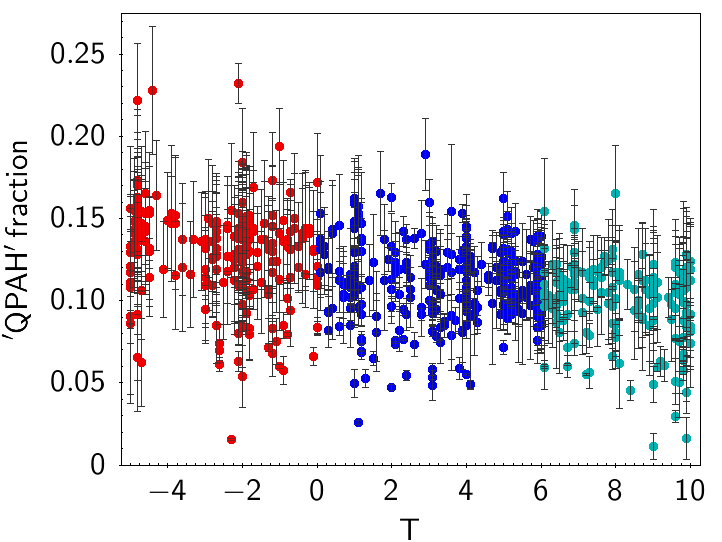}
      \caption{{\it Upper panels:} Distribution functions of the derived properties for the ensemble (black histogram) and per galaxy type (coloured histograms). 
        {Lower panels: } Distribution of derived properties as a function of galaxy type.
        In all panels, dust masses are shown in the left panels, average ISRF intensities are shown in the middle panels, and `QPAH' fraction in the right panels.} 
      \label{sl_figure2}%
 \end{figure*}
 %
 The upper panels of Fig.~\ref{sl_figure2} show the distribution functions of the model output parameters, that is, the dust mass, M$_{dust}$, the average ISRF intensity, U$_{av}$, and the `QPAH' fraction, from left to right. The distribution functions for the whole sample (753 galaxies) are shown with the black histograms, while the median values of the distributions, as well as the median absolute deviation, are listed in
%
 \begin{table}
   \tiny
      \begin{minipage}[t]{\columnwidth}
      \caption[]{Median values of M$_{dust}$, {\em U$_{av}$}, and QPAH, for the ensemble of galaxies and per galaxy type. The quoted numbers in the parentheses denote the median absolute deviation.}
      \label{sl_table4} 
      \renewcommand{\footnoterule}{}
      \begin{tabular}{ccccc}
\hline\hline
Property                              &All        &ETGs               &LTGs                  &Irrs            \\ 
\hline\hline
M$_{dust}$(10$^{6}$M$_{\sun}$)         &2.00(1.91)     &0.28(0.22)      &7.28(5.30)           &1.48(1.33)      \\
U$_{av}$\,(U$_{\sun}$)                &2.58(1.35)     &2.89(1.43)      &3.09(1.63)           &1.42(0.60) \\
`QPAH'                             &0.12(0.02)     &0.14(0.02)      &0.12(0.02)            &0.10(0.01)  \\
\hline
\end{tabular} 
\end{minipage}
\end{table}
%
 Table~\ref{sl_table4}. In the same table, we also list the median and median absolute deviation of the derived properties according to galaxy type, of which the distribution functions are shown in Fig.~\ref{sl_figure2}. The range for the M$_{dust}$ is from 100 to 1.6$\times$10$^{8}$M$\sun$, for the U$_{av}$ is from 0.3 to 7.6$\times$10$^{4}$, for the `QPAH' fraction is from 0.011 to 0.435 (we note that the Galactic value for the THEMIS model is 0.071). One nearby dwarf galaxy, ESO351-030 or Sculptor, is part of the ensemble. The value of the model-derived dust mass is $\sim$1\,M$_{\sun}$, while its model dust-SED in the FIR/submm part of the spectrum is constrained with the PACS\,70$\mu$m and 160$\mu$m emissions. The {\it WISE}\,22$\mu$m emission is an upper limit and the derived SFR is zero, hence this galaxy is not retained in subsequent analyses, while we include the results of the model-derived and calibration-derived properties in Tables A.1 and A.2.

   The histograms show that M$_{dust}$ has different distribution when considering the different galaxy types. The ETGs have lower average dust masses as compared to the other two galaxy types, for which the distributions are indistinguishable (see also Table~\ref{sl_table4}). The `QPAH' fraction has the same distribution for each galaxy type, while ETGs tend to a larger mean value but with a large standard deviation which make their 'QPAH' fraction comparable to the other two types. The average ISRF intensities are similarly distributed, when considering the ensemble or the different galaxy types, with the Irs tending to slightly lower mean value for the U$_{av}$. The lower panels of Fig.~\ref{sl_figure2} is another way to look the distribution of the derived properties per galaxy type. Overall, and considering the uncertainties, there is no trend of the derived properties with galaxy type, as all types span the whole range of values derived for each property considered. A mild trend is seen for LTGs to have values for the dust mass higher than those of  the ETGs and/or Irs \citep[see also\,][for a similar trend seen for the dust masses of LTGs using IRAS bands]{Sauvage1994}. ETGs with values of the dust mass as high as those of the LTGs exist \citep[see also][]{Lianou2016}. Galaxy misclassification may add to the scatter seen by outlier galaxies, and we have investigated this in occasions this may occur. The four ETGs seen in the lower right panel, with `QPAH' fraction higher than $\sim$0.20 or lower than 0.05, are not subject to misclassification\footnote{These galaxies are NGC\,2768, NGC\,5128, and NGC\,1404 (high `QPAH' fraction) and NGC\,1266 (low `QPAH' fraction). The first galaxy, NGC\,2768, has no constrain in the FIR/submm and was originally retained in the sample due to {\it Spitzer} constraints, which we excluded in the fit. There are a total of 8 galaxies in our sample with no FIR/submm constraint: NGC\,147, NGC\,1460, NGC\,1481, NGC\,2768, NGC\,4636, NGC\,4696, NGC\,4762, and NGC\,5119. NGC\,5128 (Centaurus\,A) has {\it Planck} constraints and is the ETG seen in the lower left panel with the larger dust mass (M$_{dust}$=3.32$\times$10$^{7}$\,M$_{\sun}$). NGC\,1404 and NGC\,1266 have {\it Herschel} constraints.}. Crossmatching our galaxy sample with the sample of luminous infrared galaxies in the GOALS sample \citep{Armus2009}, we identify 8 galaxies in common. These are NGC\,3256, NGC\,5010, NGC\,1068, NGC\,2146, NGC\,4194, NGC\,7552, NGC\,1365. All galaxies are classified as LTGs and/or Irs, but for NGC\,5010 (SO); inspecting optical images in NED, we find this classification consistent for this galaxy.

\subsection{Relations}

 \begin{figure}
  \centering
      \includegraphics[width=7.9cm,clip]{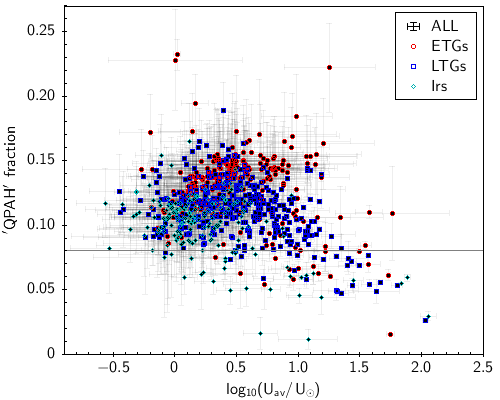}
      \includegraphics[width=7.5cm,clip]{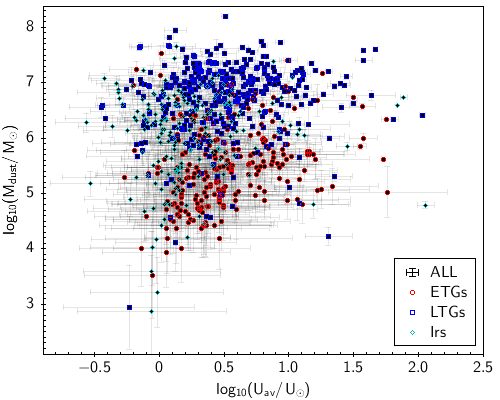}
       \caption{{\it Upper panel:} `QPAH' as a function of average ISRF intensity.
         {\it Lower panel:} M$_{dust}$ as a function of average ISRF intensity, U$_{av}$.
        }
      \label{sl_figure3}%
 \end{figure}
%
We explore the relation among the model-derived properties, that is, M$_{dust}$, U$_{av}$, and `QPAH' fraction, while Figure C.1 in the Appendix C shows the dust-specific SFR (see Section 6.1 for definition) as a function of U$_{av}$.  Fig.~\ref{sl_figure3} shows the relation between `QPAH' fraction and U$_{av}$, in the upper panel, and the relation between M$_{dust}$ and U$_{av}$ in the lower panel. A large value for the U$_{av}$ indicates an environment exposed to an, on average, most intense photon field and with high temperature.

A relation for the average ISRF intensity, U$_{av}$ (or equivalent tracers of the ISRF), and `QPAH' fraction (or equivalent tracers for the aromatic features) has been previously explored with {\it Spitzer} mid-IR bands for dwarf and starburst galaxies \citep[e.g., ][and references therein]{Wu2006,Engelbracht2008}. The correlation found between the two properties reflects the destruction and/or the reprocessing of the small dust grains (in the present study, dust grains with sizes between 0.7nm to 1.5nm) when exposed to an intense ISRF. The upper panel of Fig.~\ref{sl_figure3} shows that, for values of the log$_{10}$U$_{av}$ between -0.5 to 0.6, there is no covariance between the `QPAH' fraction and U$_{av}$, with a correlation coefficient {\it r}=0.20. For values of the log$_{10}$U$_{av}$ larger than 0.6, there is a drop of the `QPAH' fraction with increasing U$_{av}$ (negative covariance), but the scatter is large enough and the correlation coefficient is only {\it r}=-0.10. This is the case whether we consider the whole galaxy sample or separated samples according to the different galaxy type.

However, we integrate over a large enough area such that many physical processes remain unresolved to allow to uncover the finest details of the variation of the `QPAH' fraction among galaxies as a function of the ISRF intensity. When we investigate the same relation on small physical scales (150pc to 300pc linear sizes), the covariance of the two properties is clearer, that is, the larger the U$_{av}$, the smaller the `QPAH' fraction \citep[][]{Lianou2018a}. Interestingly, the overall shape of the relation between the `QPAH' fraction and U$_{av}$ is reminiscent to what is observed for H$_{\rm II}$ regions in M101 by \citet[][]{Gordon2008}, where the equivalent widths of the aromatic features are plotted against the ionisation index \citep[defined in][]{Gordon2008}, favouring a power law with a constant functional form, and a shape similar to what is seen in the upper panel of Fig.~\ref{sl_figure3}.

The Galactic value for the `QPAH' fraction is marked with the solid line in the upper panel of Fig.~\ref{sl_figure3}: for a U$_{av}$=1U$_{\sun}$, the value of `QPAH' fraction is 0.071. The galaxies with U$_{av}\sim$1\,U$_{\sun}$ have a `QPAH' fraction spread over a large range of values between $\sim$0.08 to $\sim$0.23 and with $\langle$QPAH$\rangle$=0.12$\pm$0.03s. The ETGs with `QPAH' fraction value larger than 0.20 are discussed in Section 6.1. There are only a few galaxies with `QPAH' fraction consistent with the Galactic one, with most galaxies having larger `QPAH' fractions. The larger `QPAH' fraction indicates that the very small dust grains (hydrogenated amorphous carbon) exposed to a similar U$_{av}$ as the Galactic one are less affected in different galactic environments. While our Galaxy and its properties may not be representative of its type, nevertheless whether this variation reflects a change in the dust properties, i.e size distribution or composition, remains open. 

The dust mass, M$_{dust}$, versus the average ISRF intensity, U$_{av}$, is shown in the lower panel of  Fig.~\ref{sl_figure3}, where a large scatter is seen. No correlation between the two properties is expected {\em a priori}. Whether we consider the galaxy ensemble or individual galaxy types, the correlation is close to zero, while the covariance is negative for the ETGs and Irs and positive for the LTGs.  When we study the variations of the dust mass and the U$_{av}$ on small physical scales \citep[][]{Lianou2018a}, we find an anti-correlation between them, indicating that regions with larger U$_{av}$ (or dust temperatures) tend to lower values for the dust mass. For the global scale analyses of the galaxy ensemble, as we are integrating fluxes over a large area for each galaxy, regions with a variety of physical conditions present (temperatures or radiation field intensities, dust grain properties) are averaged, adding to the scatter seen in Fig.~\ref{sl_figure3}. 

\section{Star formation and dust-SED-derived properties}
 
\subsection{Stellar masses and star formation rates}

 \begin{figure}
  \centering
      \includegraphics[width=7.9cm,clip]{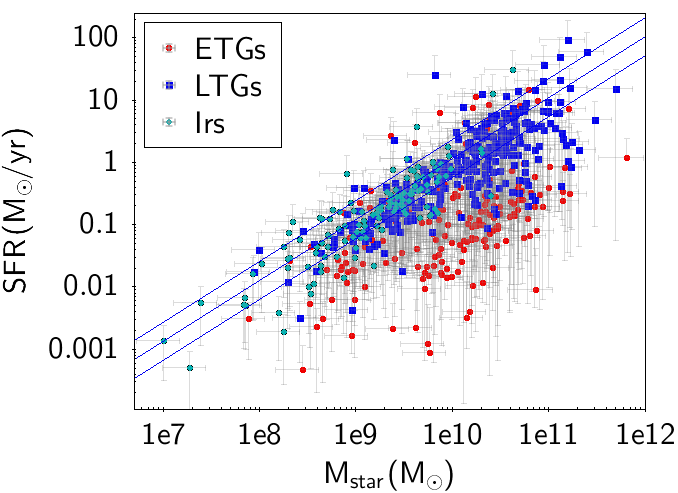}
      \includegraphics[width=7.5cm,clip]{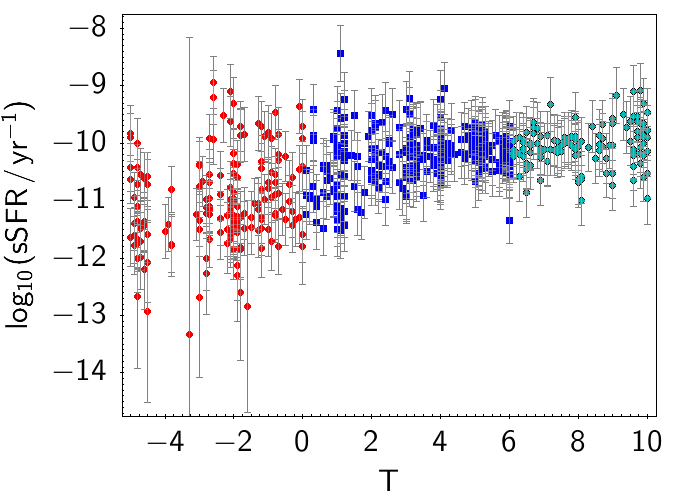}
      \includegraphics[width=7.5cm,clip]{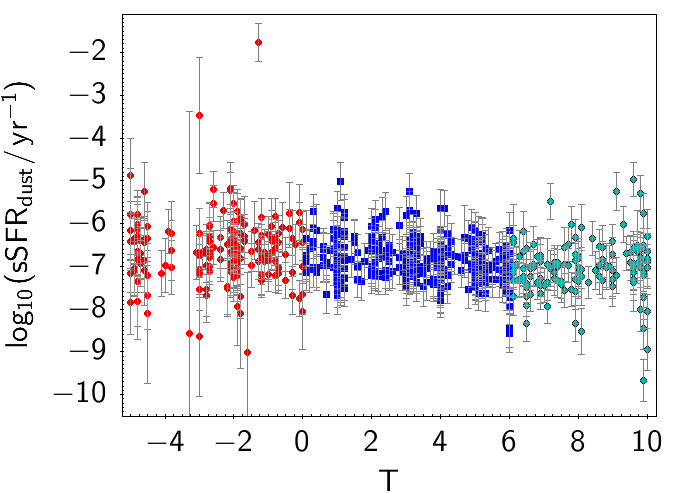}
      \caption{{\it Upper panel:} SFR as a function of the stellar mass. The blue line shows a fit to the LTGs and Irs, along with the lines with $\pm$0.3\,dex scatter in the y-axis. 
        {\it Middle panel:} Specific star formation rate, sSFR, versus galaxy type. 
        {\it Lower panel:} Dust-specific star formation rate, sSFR$_{dust}$, versus galaxy type. }
      \label{sl_figure4}
 \end{figure}
%
 The upper panel in Fig.~\ref{sl_figure4} shows the SFR as a function of the stellar mass, M$_{\star}$, for the 753 galaxies studied here. The ETGs lie in a sequence below the LTGs, while some ETGs are scattered above and below the sequence defined by the LTGs \citep[see also][]{Lianou2016}. Star formation in ETGs can have multiple origins: from rejuvenation of the formation of stars through the accretion of gas-rich satellites to the suppression of star formation in a post-starburst event. These processes would place ETGs anywhere above the sequence defined by the bulk of the ETGs. The Irs, on the other hand, are found to lie in the LTGs sequence. An error-weighted linear least squares power-law fit to the LTGs and Irs is shown with the blue line, and the equation describing the fit is given as: SFR = 1.89$\times$10$^{-10}$ $\times$ M$_{\star}^{0.98}$ (M$_{\sun}$/yr), with a correlation coefficient {\it r}$\sim$0.9. This slope is consistent with the slope of $\sim$one found to hold for this relation for galaxies independent of redshift \citep{Elbaz2011}.

The middle panel in Fig.~\ref{sl_figure4} shows the ratio present-to-past star formation encoded in the specific SFR (sSFR = SFR / M$_{\star}$), as a function of Hubble type. The SFR (WISE 22$\mu$m) is a tracer of the present star formation (after correction for the contribution of the evolved stellar populations to the mid-IR emission), while the M$_{\star}$ (WISE 3.4$\mu$m) is a tracer of the integrated previous star formation, with low-mass stars the dominant population \citep[with an average age of $\sim$10Gyr, e.g.,][]{Madau2014}. \citet{Jarrett2013} discuss a similar relation and note the increase of the gas supply, encoded in the SFR, as the Hubble type increases. In Fig.~\ref{sl_figure4}, there is a similar trend of increasing sSFR when the Hubble type increases, possibly reflecting the higher gas supply and star formation occurring in irregular galaxies. ETGs show a larger scatter as compared to LTGs and Irs, reflecting their complex star formation and ISM histories of their evolution \citep{Lianou2016}. The trend of the sSFR with galaxy type is very weak.

If we look at the dust-specific SFR (defined as sSFR$_{dust}$=SFR/M$_{dust}$) as a function of Hubble type (lower panel of Fig.~\ref{sl_figure4}), there is no trend seen, overall, while few ETGs seem to be scattered. Here, the dust mass is related to the gas mass (assuming a gas-to-dust mass ratio relation), hence the dust-specific SFR is a proxy for the star formation efficiency (SFE = SFR/Mgas). This is almost constant in our local galaxy sample. There are two ETGs that scatter towards higher dust-specific SFRs. These are ESO\,434-040 and NGC\,2110. Both galaxies are classified as lenticulars (T=-1.3 and -3, respectively) and after inspecting optical images, we find that their morphology is consistent with the assigned morphology. The SFRs for these two galaxies are 14\,M$_{\sun}\,yr^{-1}$ and 7\,M$_{\sun}\,yr^{-1}$, respectively, while the dust masses are 819\,M$_{\sun}$ and 2.1$\times$10$^{4}$M$_{\sun}$, respectively. An elevated SFR, due to an elevated {\it WISE} 22$\mu$m emission as compared to the FIR/submm, combined with the small amount of dust mass leads to their outlier nature in the bottom panel of  Fig.~\ref{sl_figure4}.

\subsection{Dust versus SFR and stellar mass}

 \begin{figure}
  \centering
      \includegraphics[width=7.9cm,clip]{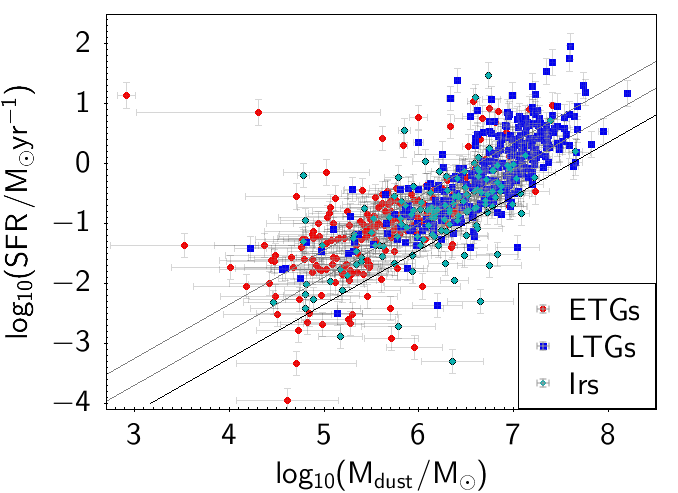}
      \includegraphics[width=7.9cm,clip]{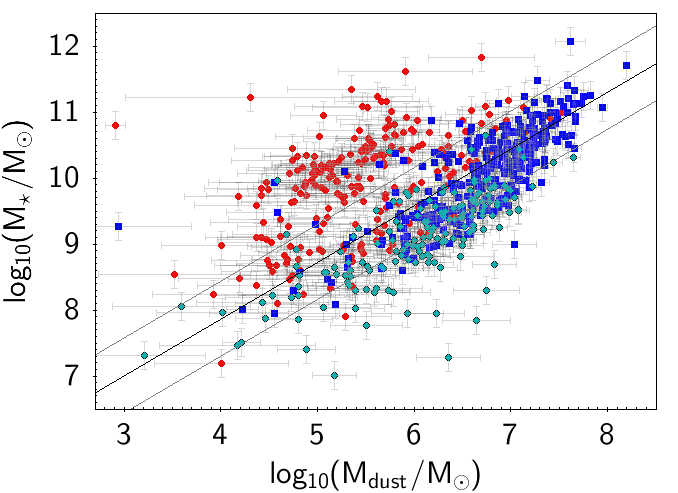}
      \caption{SFR versus M$_{dust}$ (upper panel), and M$_{\star}$ with M$_{dust}$ (lower panel). ETGs are shown with red symbols, late-type galaxies are shown with blue symbols, and irregulars are shown with light blue symbols. The error bars represent the standard deviation for the M$_{dust}$, and the propagated uncertainty (photometric and calibration) for the SFR and M$_{\star}$. The black solid line in the upper panel is the relation found by \citet{daCunha10b} along with the intrinsic scatter of the relation. The black solid line in the lower panel is the relation between M$_{dust}$ and M$_{\star}$, along with the $\pm$1-$\sigma$ intrinsic scatter.  }
      \label{sl_figure5}
 \end{figure}
%
 Fig.~\ref{sl_figure5} shows the relation between M$_{dust}$ and SFR (upper panel) and the M$_{\star}$ (lower panel), for the galaxy sample studied here. There is a trend in our sample between M$_{dust}$ and SFR. Such a tight relation between M$_{dust}$ and SFR was also found by \citet{daCunha10b} in a sample of low-redshift galaxies drawn from SDSS-DR6. Our galaxy sample falls within the relation of \citet{daCunha10b}, where the majority of the scatter is due to ETGs. If M$_{dust}$ traces the gas, assuming a gas-to-dust mass ratio (GDR), then the relation between the SFR and M$_{dust}$ reflects another representation of the KS law, which we investigate in Section 6.3. 

 In the lower panel of Fig.~\ref{sl_figure5}, we show the relation between M$_{\star}$ and M$_{dust}$, where the LTGs and Irs and ETGs are separated. This reflects the lower dust masses associated with ETGs at the same stellar mass, compared to that of LTGs.
The tightness between M$_{dust}$ and M$_{\star}$ for the LTGs is remarkable, and a error-weighted linear fit yields:
 \begin{equation}
 log_{10} (M_{\star}) = 0.86 \times \log_{10}(M_{dust}) + 4.42 ,
 \end{equation}
and the 1-$\sigma$ intrinsic scatter of this relation is 0.66\,dex in M$_{dust}$ and the correlation coefficient {\it r}$\sim$0.8.

Dust contains a significant fraction of the metals in the ISM and the mass of dust observed in a region is related to both the ISRF intensity and metallicity \citep{Dwek1998}. Assuming that M$_{\star}$, that is, the average stellar population, is a measure of the metals within a galaxy \citep{Tremonti2004}, then the relation between M$_{\star}$ and M$_{dust}$ reflects the relation between metallicity and M$_{dust}$. While M$_{\star}$ and metallicity show a tight relation with a scatter of $\pm$0.1\,dex, the latter is much smaller than the intrinsic scatter of M$_{\star}$ and M$_{dust}$ we see in Fig.~\ref{sl_figure5}. This suggests a second parameter to explain the scatter in Fig.~\ref{sl_figure5}. The relation between M$_{dust}$ and metallicity shows a larger scatter \citep[$\sim$0.37dex,][]{Remy14,Lisenfeld1998}, making metallicity an important parameter to the scatter seen in the lower panel of Fig.~\ref{sl_figure5}. In addition, the scatter seen between U$_{av}$ and M$_{dust}$ (see Fig.~\ref{sl_figure3}) means that U$_{av}$ plays a role in the scatter seen in Fig.~\ref{sl_figure5}. Whether the fundamental property driving the relation in Fig.~\ref{sl_figure5} is  M$_{\star}$, and how metallicity and U$_{av}$ connect to the scatter seen therein, will be investigated in a subsequent study. To perform this investigation, we need spatially resolved studies. 

 \subsection{Extended KS law, gas mass, and the GDR}

Star formation in a galaxy depends on the available gas supply \citep{Kennicutt1998} through a rather complex process \citep{Lada2013}. The link among gas, dust, and stellar mass is expressed through the extended KS law, which holds both within and among galaxies \citep{Rahmani2016,Shi2011}. In the following, we investigate the KS law by making various assumptions about the GDR to derive the gas mass from the dust mass (Section 6.3.1). Then we use the extended KS law to derive an approximate gas mass and GDR, knowing the SFR and M$_{\star}$, without making an assumption about the GDR (Section 6.3.2).

\subsubsection{KS law assuming a constant GDR}

 \begin{figure}
  \centering
  \includegraphics[width=7.9cm,clip]{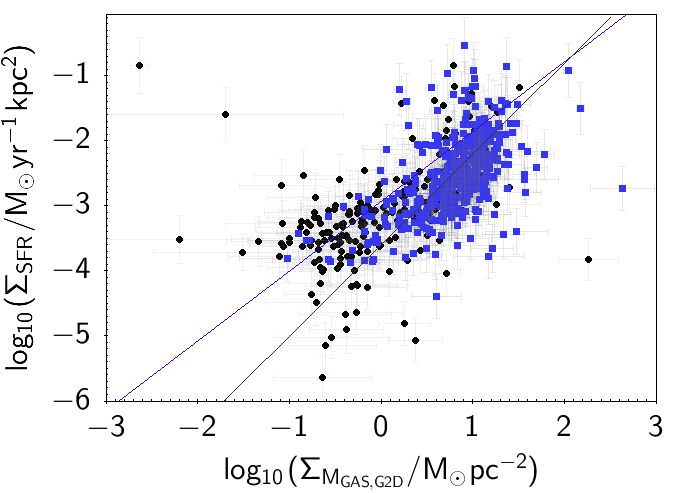}
  \includegraphics[width=7.9cm,clip]{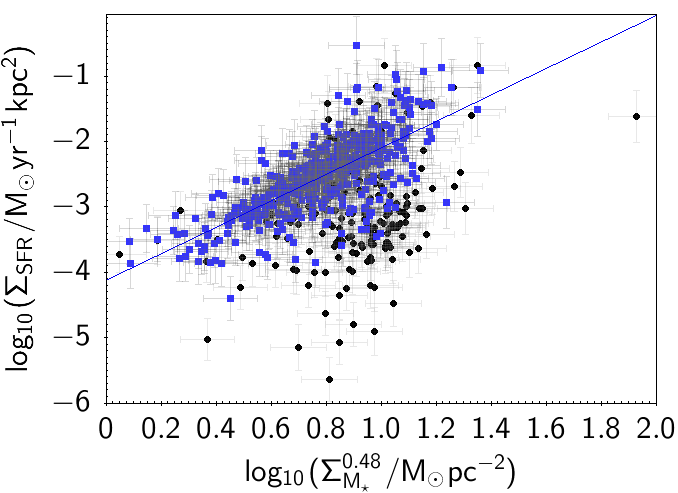}
  \caption{{\it Upper panel}: Surface density of SFR versus surface density of gas mass, derived assuming a constant GDR$\sim$270. The grey solid line represents the KS law, and the solid blue line is the fit we find to the LTGs and Irs (marked with blue symbols).
    {\it Lower panel}: Surface density of SFR versus surface density of stellar mass, i.e., the extended KS law. The solid blue line is the fit we find to the LTGs and Irs (marked with blue symbols). }
      \label{sl_figure6}%
 \end{figure}
%
 The upper panel of Fig.~\ref{sl_figure6} shows the relation between the surface densities of SFR and gas mass, $\Sigma_{SFR}$ and $\Sigma_{M_{GAS,GDR}}$, respectively, that is, the KS law. The gas mass, M$_{GAS,GDR}$, is derived from the dust mass assuming a constant GDR equal to $\sim$270 \citep[mean value adopted from][these authors study an assembled galaxy sample with a skewed non-unimodal distribution in stellar mass]{Remy14}. Assuming another GDR value \citep[e.g., GDR$\sim$72.4 from][or any other constant value]{Sandstrom13} means that the gas masses will vertically shift, leaving the slope of the KS relation unaffected. An error-weighted linear fit to the LTGs and Irs' $\Sigma_{SFR}$ and $\Sigma_{M_{GAS,GDR}}$ yields the following relation:
 \begin{equation}
 log_{10} (\Sigma_{SFR}) = (1.1\pm 0.1) \times log_{10}(\Sigma_{M_{GAS,GDR}}) - (2.9\pm 0.2) .
 \end{equation} 
The slope of the LTGs- and Irs-fitted KS law (solid blue line) is shallower than the slope of the KS law \citep[equal to 1.4; light grey line][]{Kennicutt1998}. The difference in the slopes is expected for two reasons. First, there is the assumption of a single constant GDR applied to all galaxies, which means that gas and dust are well mixed at all physical scales. This may not hold in regions and/or galaxies of low metal abundance \citep[][and references therein]{Draine2007,Berta2016,Tacconi2018}. The low metal abundance also means that there is gas not traced by the dust \citep[][and references therein]{Bolatto2013}. Second, the dependence of the stars to the KS law was neglected. Both metallicity and stellar surface density will impact the derived slope of the above equation.

\subsubsection{Extended KS law to derive the GDR} 

We use the results of the global analysis of \citet{Shi2011} to derive the gas mass and GDR for our galaxy sample. The extended KS law is derived from the relation between the surface densities of stellar mass and star formation efficiency, SFE, that is, the ratio of the surface densities of SFR and M$_{GAS}$, i.e.:
\begin{equation}
  SFE = \Sigma_{SFR} / \Sigma_{M_{GAS}} ,
\end{equation} 
where the surface densities of SFR and M$_{GAS}$ have units  M$_{\sun}$\,yr$^{-1}$\,kpc$^{-2}$ and M$_{\odot}$\,kpc$^{-2}$, respectively. The relation between SFE and $\Sigma_{M_{\star}}$ is \citep[equation 6 in ][]{Shi2011}: 
\begin{equation}
 SFE = 10^{-10.28\pm 0.08} \times \Sigma_{M_{\star}}^{0.48\pm0.04} , 
 \end{equation}
where SFE is expressed in yr$^{-1}$ and $\Sigma_{M_{\star}}$ in M$_{\odot}$\,pc$^{-2}$. In eq.~5, SFE does not depend on $\Sigma_{M_{GAS}}$, while studies on small physical scales find a dependence on both $\Sigma_{M_{GAS}}$ and $\Sigma_{M_{\star}}$ \citep{Shi2011,Rahmani2016}. 

From eqs.~4 and 5, we derive the following expression: 
\begin{equation}
 \Sigma_{SFR} = (10^{-10.28\pm 0.08} \times \Sigma_{M_{GAS}}) \times \Sigma_{M_{\star}}^{0.48\pm0.04}.
 \end{equation} 
Eq.~6 gives an alternative way to derive an estimate of M$_{GAS}$, knowing the surface densities of SFR and M$_{\star}$, most useful to higher redshift galaxies \citep{Scoville2017,Scoville2014}. We show the relation between $\Sigma_{SFR}$ and $\Sigma_{M_{\star}}^{0.48}$ in the lower panel of Fig.~\ref{sl_figure6}. The error-weighted linear fit to only the LTGs and Irs (blue solid line) yields the following:
 \begin{equation} \label{eq:Mgas}
 log_{10} (\Sigma_{SFR}) = (2.02\pm 0.04) \times log_{10}(\Sigma_{M_{\star}}^{0.48\pm0.04}) - (4.12\pm 0.08) .
 \end{equation}

 \begin{figure}
  \centering
  \includegraphics[width=8cm,clip]{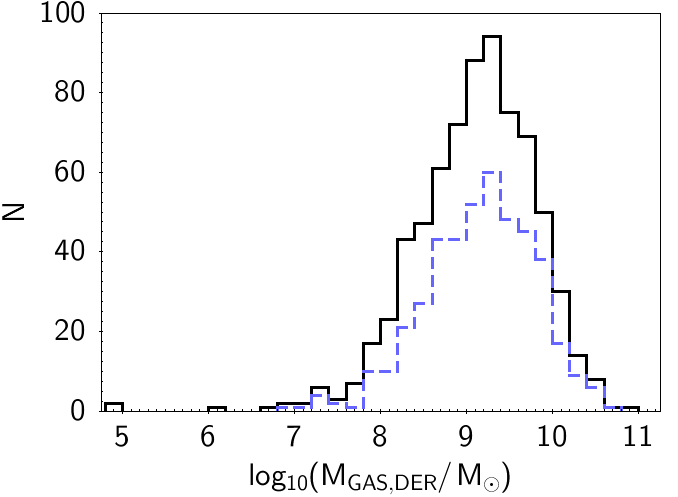}
  \includegraphics[width=8cm,clip]{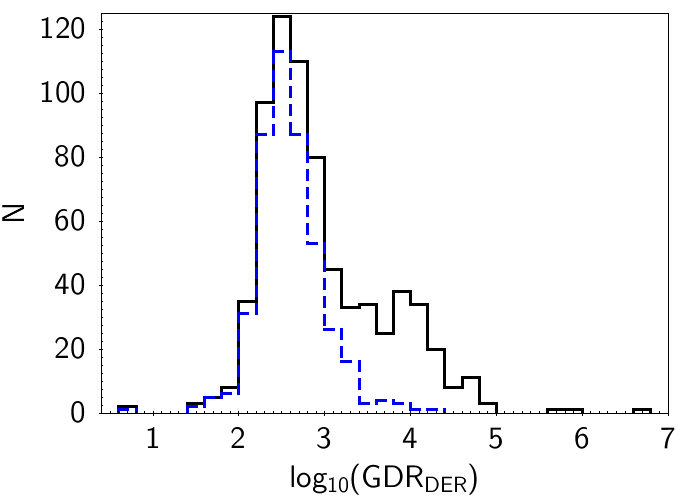}
  \caption{{\it Upper panel}: Distribution function of the derived gas mass, M$_{GAS,DER}$, assuming an extended KS law to hold. The blue dotted histogram shows the M$_{GAS,DER}$ for the LTGs and Irs, while the black solid line shows the derived M$_{GAS,DER}$ for all galaxies, assuming that the ETGs also follow the same equation~\ref{eq:Mgas}, defined by the LTGs and Irs.  \, 
    {\it Lower panel}: Distribution function of the GDR$_{DER}$, assuming an extended KS law to hold. The dotted blue and solid black histograms are defined in the same way as in the upper panel. }
      \label{sl_figure7}%
 \end{figure}
%
 Eqs.~6 and 7 allows us to derive the gas mass, M$_{GAS,DER}$, and its distribution is shown in the upper panel of Fig.~\ref{sl_figure7}. The distribution for the LTGs and Irs is the blue dotted histogram, and the distribution for the ensemble is the black solid line assuming that ETGs follow the same eq.~\ref{eq:Mgas}. The mean and standard deviation of M$_{GAS,DER}$  for the LTGs and Irs is 3.44($\pm$6.05)$\times$10$^{9}$\,M$_{\sun}$, and for the ensemble is 3.35($\pm$6.04)$\times$10$^{9}$\,M$_{\sun}$. \citet{Leroy08} note that the connection of the SFE to stellar mass is appropriate for systems which are HI-dominated, hence the gas mass we derive might be related more to HI gas. We remind that in the case of the ETGs of our sample these are selected to have {\it Herschel} FIR/submm emission. Many ETGs are gas- and dust-poor and the ETGs of our sample are not representative of the broader population of ETGs.

Comparing with the KS law, and considering only LTGs and Irs, the average ratio of the gas mass derived from the KS law, M$_{GAS,KS}$, over the gas mass derived from the extended KS law, M$_{GAS,DER}$, is 0.82$\pm$0.85, with a median value of 0.64. The same ratio becomes 0.71$\pm$0.85, with a median value of 0.54, when we include the ETGs. The dependence of SFE on either the gas mass or the stellar mass is described by similar power law indexes (0.4 and 0.48, respectively, for the KS and extended KS law), then it is not surprising that the two derivations are in agreement, on average.  

The lower panel of Fig.~\ref{sl_figure7} shows the distribution of the GDR, GDR$_{DER}$, where the gas mass is estimated from the extended KS law (eq.~7 above). The histogram of the GDR$_{DER}$ is skewed due to the dust mass and the median GDR$_{DER}$ is 370 for the LTGs and Irs (blue dotted line), while the median GDR$_{DER}$ is 550 for the ensemble (black solid line). The GDR$_{DER}$ for the LTGs and Irs is consistent with studies of LTGs and Irs. The median GDR$_{DER}$ of the whole galaxy sample is larger due to the combination of a smaller dust mass content and larger galaxy masses. 

Considering only ETGs, the median GDR$_{DER}$ is 5462, and a median M$_{GAS,DER}$ of 1.5$\times$10$^{9}$\,M$_{\sun}$; the median values for the GDR$_{DER}$ and M$_{GAS,DER}$ we estimate here originate from an `extrapolation' of eq.~\ref{eq:Mgas}. With a median M$_{\star}$ of 1.1$\times$10$^{10}$\,M$_{\sun}$, the GDR$_{DER}$ for the ETGs is towards the higher end of the observationally derived GDR in \citet[][lower panel of their figure 7; for similar stellar masses, the GDR ranges from $\sim$300 to $\sim$3500]{Lianou2016}. The mean observed GDR is 631, and the range is between 200 to 2000. ETGs have higher stellar masses as compared to their global ISM mass budget and an extended KS law should be even more relevant, yielding a steeper slope in eqs.~5 and \ref{eq:Mgas}, to accommodate their poorer ISM content.

 \section{Summary and conclusions}

One of the largest datasets of local galaxies coupled with an hierarchical Bayesian dust-SED model has allowed us to obtain statistical constraints on their dust emission and star formation properties. The sample consists of 753 galaxies and forms a sub-sample of the DustPedia galaxies \citep{Clark2018Dust}, a catalogue of photometric measurements of $\sim$850 galaxies homogeneously treated. The dust-SED model incorporates the THEMIS dust grain properties and allows us to treat the galaxy sample as an ensemble. We derive the model dust-SEDs and physical properties, that is, the average ISRF intensity (U$_{av}$), the dust mass (M$_{dust}$), and the mass fraction of very small grains (`QPAH' fraction; very small hydrogenated amorphous carbon grains with sizes between 0.7nm and 1.5nm).
While we examine the relation among the dust-SED derived properties, we also examine their relation to the star formation rate (SFR; from WISE 22$\mu$m, corrected for the evolved star contribution \citep{Temi2009,Davis2014}) and stellar mass (M$_{\star}$; from WISE 3.4$\mu$m). We explore how we can use dust masses and the star formation rates and masses to estimate the amount of gas mass present in the galaxies under study. Our findings can be summarised as follows.

First, the model-derived dust-SEDs of each galaxy, and the mean model dust-SED, are shown in Fig.~\ref{sl_figure1b}, where ETGs on average emit less power across the spectrum than LTGs or Irs. There are variations of the mid-IR spectral region of the model dust-SED per galaxy type, and the mean model dust-SED of the ensemble better approximates that of the Irs in the mid-IR part (Fig.~\ref{sl_figA2} and  Fig.~\ref{sl_figure1}).

Second, as expected, the distribution of the dust masses for the ETGs is different than that of the LTGs or Irs, while the U$_{av}$ and `QPAH' fraction do not show any dependence on galaxy type (Fig.~\ref{sl_figure2}).

Third, there is no clear relation among the different properties derived from the dust-SED modelling, which may hint at averaging effects on global galaxy scale measurements (see Fig.~\ref{sl_figure3}).
What is interesting is the variation of the `QPAH' fraction for galaxies with a value of U$_{av}$ similar to the Galactic value, as they deviate from the Galactic `QPAH' fraction (0.071). This may hint at a variation, or evolution, of the dust grain properties, either composition and/or size distribution, related to galaxy environment, which deserves further examination on smaller physical scales.

Fourth, the ensemble is characterised by a tight relation between dust mass and stellar mass for the LTGs and Irs (see Fig.~\ref{sl_figure5}). The relation between dust mass and stellar mass shows a large scatter, with metallicity and U$_{av}$ likely being the second parameters.   

Last, we use the extended KS law to derive an estimate of the gas mass, and we compare this to the gas mass derived both from the dust assuming a constant GDR and from the KS law (Fig.~\ref{sl_figure6}). The derived gas mass from the extended KS law is on average $\sim$20\% higher than that derived from the KS law, and a large standard deviation indicates the importance of the average star formation present to regulate star formation and gas supply \citep{Dib2017}. The importance of the average star formation should become more relevant in the case of very massive and ISM--poor galaxies. Assuming that the extended KS law well describes our galaxy sample, we have inferred a GDR with an average of 370 (or 550 when we include the ETGs). 

We emphasise that the strength of our results rely both on one of the largest galaxy sample in the local Universe with imaging treated homogeneously and on the statistical constraints posed on the properties of the galaxy sample modelled as an ensemble with a hierarchical Bayesian dust-SED model. These two elements make this study unique. As a follow up to what is presented here, we perform small physical scale analyses of as many galaxies as permitted (spatial constraints), so as to be able to understand the local variations of the dust properties versus the star formation history \citep[][Lianou et al.~in preparation]{Lianou2018a}.  
      
\begin{acknowledgements}
  We are grateful to an anonymous referee for a careful reading of the manuscript and useful suggestions that helped us to improve its content. SL thanks Kostas Themelis for discussions about hierarchical Bayesian modelling techniques, Elisabetta Micelotta for discussions about dust grain models, and David Elbaz for discussions on stellar masses. Chris Clark and Frederic Galliano are acknowledged for clarifications on the photometric measurements and on the dust-SED model, respectively. We thank Frederic Galliano for his dust-SED code. SL and this work have benefited from science discussions with Marc Sauvage. 
        Support for this  work has been provided by DustPedia, a collaborative focused research project supported by the European Union under the Seventh Framework Programme (2007-2013) call (proposal no. 606847), with participating institutions: Cardiff University, UK; National Observatory of Athens, Greece; Ghent University, Belgium; Universit\'{e} Paris Sud, France; National Institute for Astrophysics, Italy and CEA (Paris), France. SL wishes to thank the members of the staff at the Institut d'Astrophysique de Paris \& at the National \& Kapodistrian University of Athens for their hospitality and support during which portions of this work were completed.
\\
   This research made use of several facilities and open source software: Astropy \citep[\url{http://www.astropy.org}]{Astropy13}, a community-developed core Python package for Astronomy; NASA/IPAC Extragalactic Database (NED), operated by the Jet Propulsion Laboratory, California Institute of Technology, under contract with the National Aeronautics and Space Administration; NASA/ IPAC Infrared Science Archive (IRSA), operated by the Jet Propulsion Laboratory, California Institute of Technology, under contract with the National Aeronautics and Space Administration; NASA's Astrophysics Data System Bibliographic Services.
\end{acknowledgements}

\bibliographystyle{aa}
\bibliography{biblio}{}

\begin{appendix}

\section{Model and observed dust-SEDs for example galaxies, and per Hubble type}

 \begin{figure}[th!]
  \centering
      \includegraphics[width=8cm,clip]{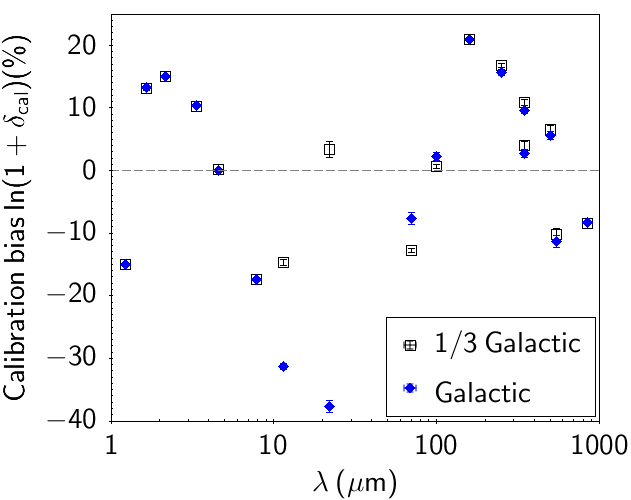}
      \caption{Comparison between the model-derived and the observed photometric data points for two different assumptions of the abundance of the small grains (see Section 3). The open black squares correspond to an abundance of one-third the Galactic value, and the solid blue diamonds correspond to an abundance of small grains equal to the Galactic one. }
      \label{sl_figA1}%
 \end{figure}
%
 Fig.~\ref{sl_figA1} shows the model-derived versus the observed photometric measurement data points per wavelength, for two different choices of the abundance of the small grains (see Section 3). The open black squares correspond to the abundance used in the main text (one third the Galactic value), while the solid blue diamonds correspond to an assumption of the abundance of the small grains equal to the Galactic one. This figure demonstrates that the mid-IR bands are better constrained in the former case. A difference of less than $\sim$20\% is seen for the observed-modelled photometric points for the chosen abundance of the small grains (one third Galactic). This difference is valid for all MCMC iterations, as evidenced by the small standard deviations relative to the deviation from the x-axis.
 A similar investigation was performed to validate the performance of the modelling when we do not use the {\it Planck} bands.
 
 \begin{figure}[th!]
  \centering
      \includegraphics[width=6.7cm,clip]{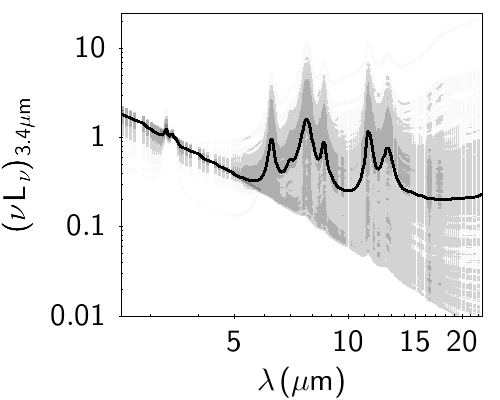}
      \includegraphics[width=6.7cm,clip]{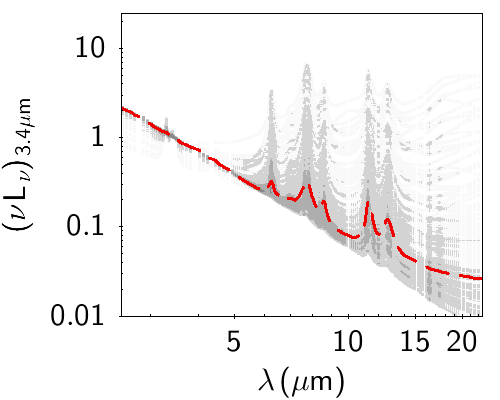}
      \includegraphics[width=6.7cm,clip]{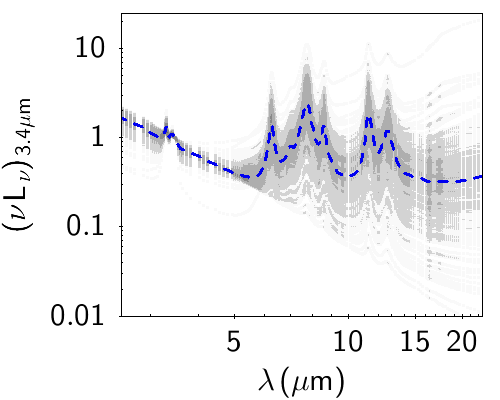}
      \includegraphics[width=6.7cm,clip]{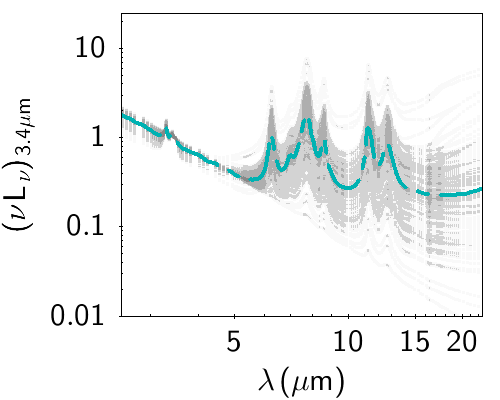}
      \caption{Exact same as in Fig.~\ref{sl_figure1b}, but zooming into the mid-IR region of the spectrum. The upper panel corresponds to the model dust-SEDs of individual galaxies, the middle upper panel to the ETGs, the middle lower panel to the LTGs, and the lower panel to the Irs. The mid-IR part of the spectrum ranges from  2.5$\mu$m to 22.5$\mu$m. }
      \label{sl_figA2}%
 \end{figure}
 %
Fig.~\ref{sl_figA2} shows a zoom-in of the mid-IR spectral region of the model dust-SEDs of Fig.~\ref{sl_figure1b}. 
The full sample of the galaxies, which were modelled as an ensemble, is shown in Table~\ref{sl_tableA1} (only the first 45 entries of the galaxy sample in Table~\ref{sl_tableA1} is shown here, while the full list of the galaxies is available in the online version of the journal), along with the model-derived and the calibration-derived properties. The columns show:
   (1) the galaxy name;
   (2) and (3) the equatorial coordinates, RA and Dec in degrees;
   (4) the revised Hubble type, T;
   (5) the major axis diameter, D25 in arcmin, at which the optical surface brightness falls beneath 25~mag/arcsec$^{2}$ \citep[adopted from][]{Clark2018Dust};
   (6) the distance, in Mpc, corresponding to the best distance in \citet{Clark2018Dust};
   (7) the model-derived dust mass, M$_{dust}$ in M$_{\sun}$ (see Section 3);
   (8) the standard deviation of the dust mass, $\sigma _{M_{dust}}$ in M$_{\sun}$;
   (9) the average radiation field intensity, U$_{av}$ in U$_{\sun}$;
   (10) the standard deviation of the average radiation field intensity, $\sigma _{U_{av}}$ in U$_{\sun}$;
   (11) the `QPAH' fraction;
   (12) the standard deviation of the QPAH fraction, $\sigma _{QPAH}$;
   (13) the star formation rate, SFR in M$_{\sun}$/yr (see Section~2.2); 
   (14) the uncertainty in the SFR, $\sigma _{SFR}$ in M$_{\sun}$/yr;
   (15) the stellar mass, M$_{\star}$ in M$_{\sun}$ (see Section~2.2);
   (16) the uncertainty in the stellar mass, $\sigma _{M_{\star}}$ in M$_{\sun}$.
 
 \begin{figure*}[th!]
  \centering
      \includegraphics[width=7.4cm,clip]{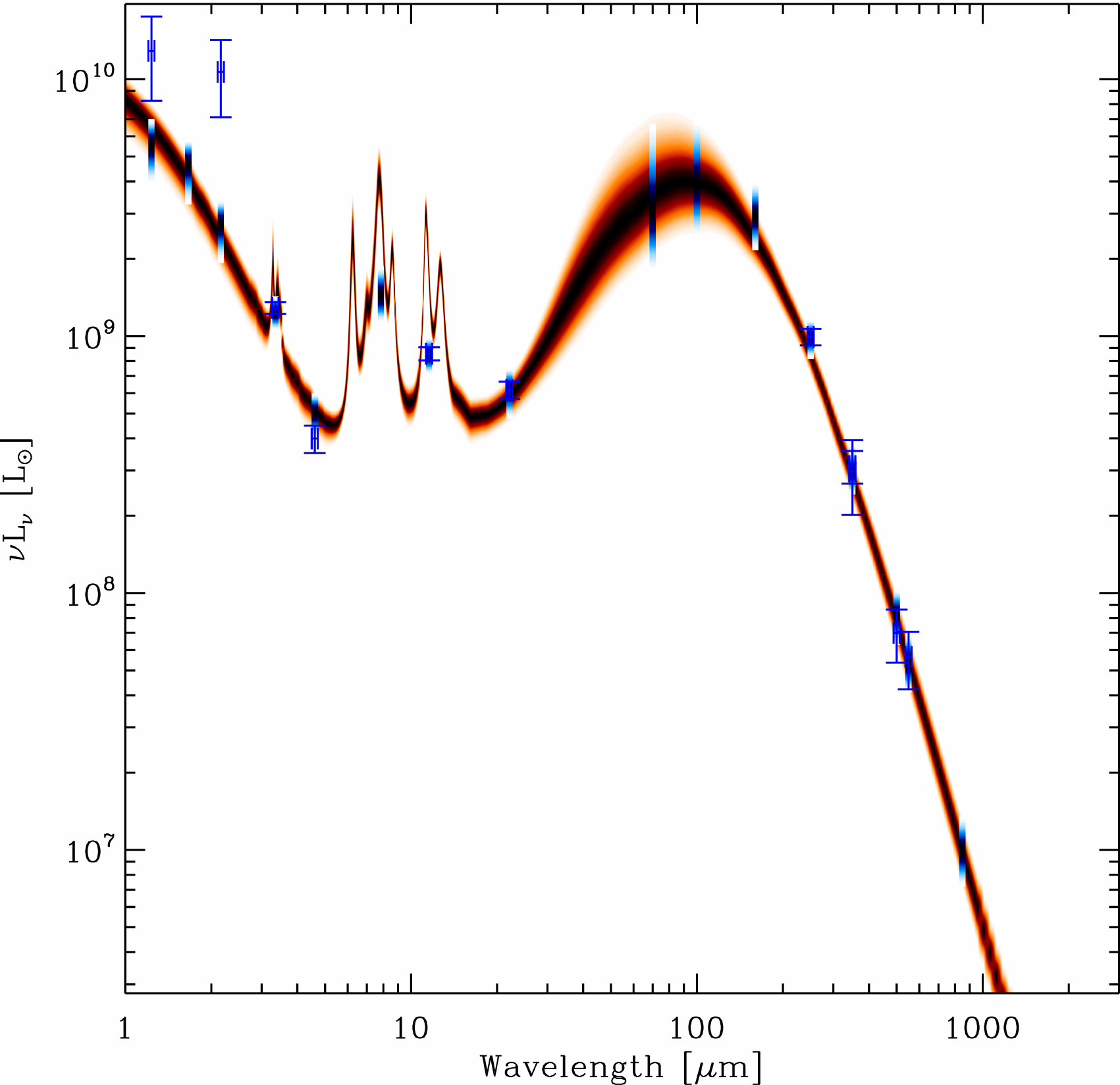}
      \includegraphics[width=7.1cm,clip]{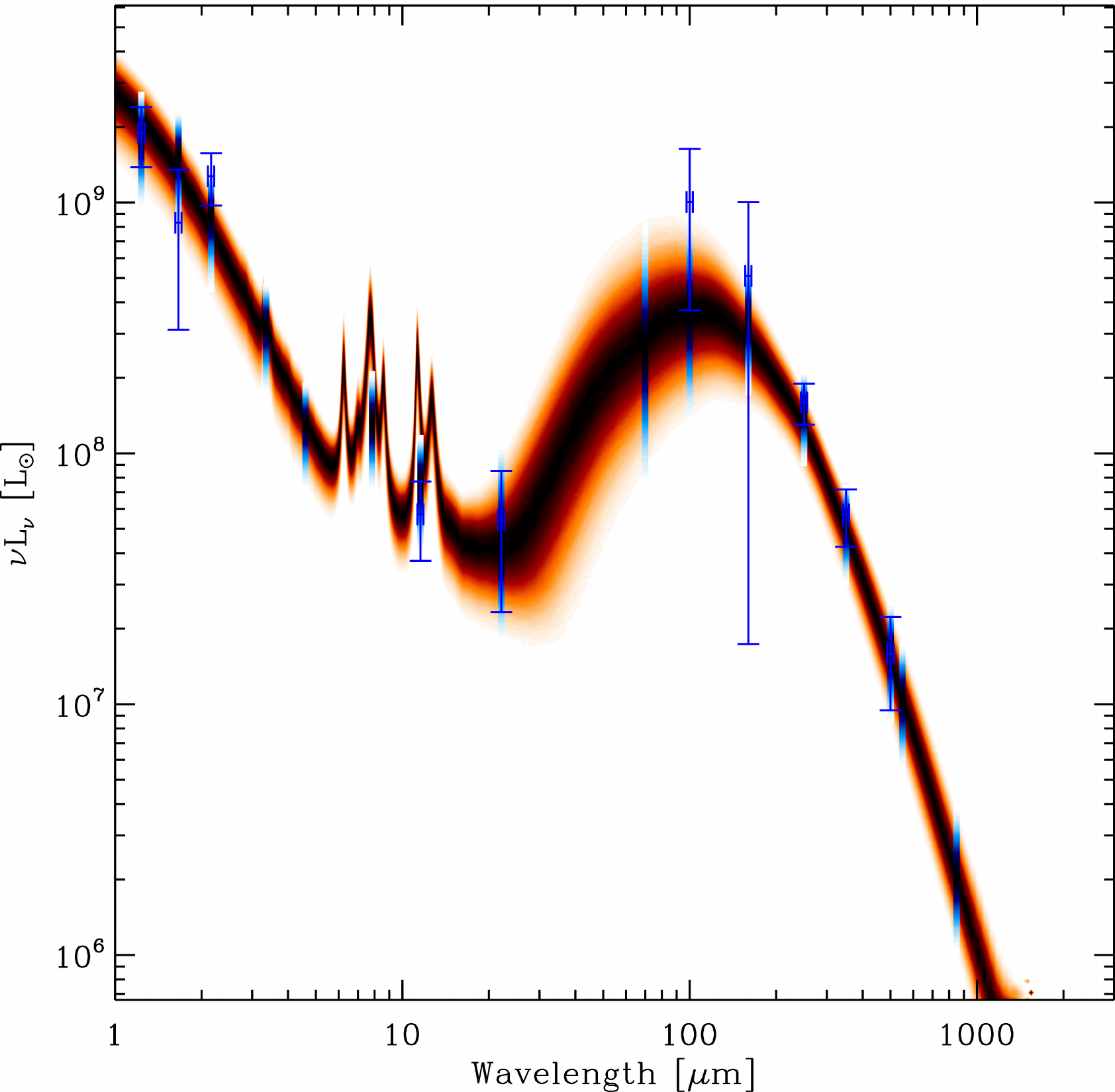}
      \includegraphics[width=7.4cm,clip]{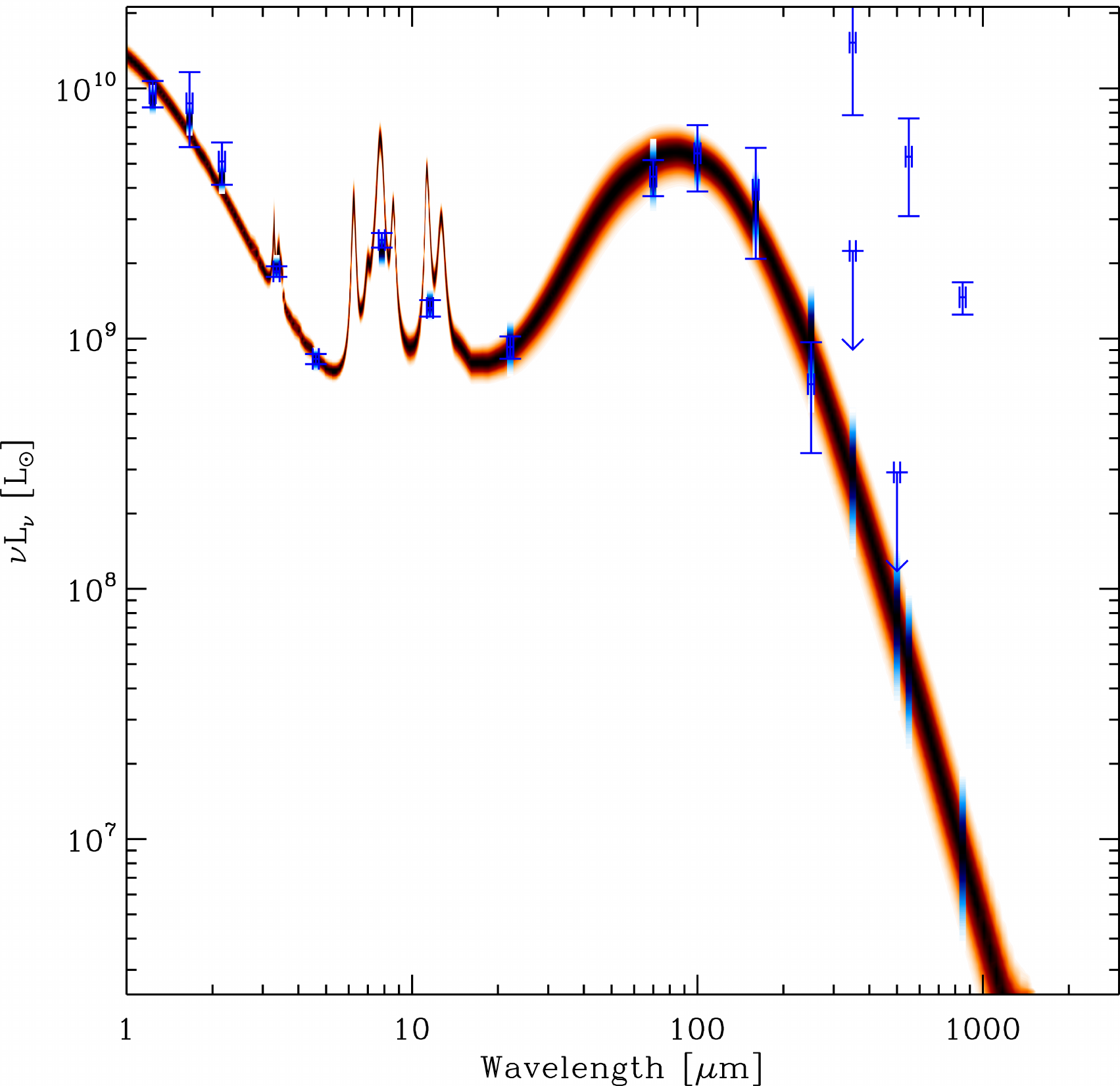}
      \includegraphics[width=7.1cm,clip]{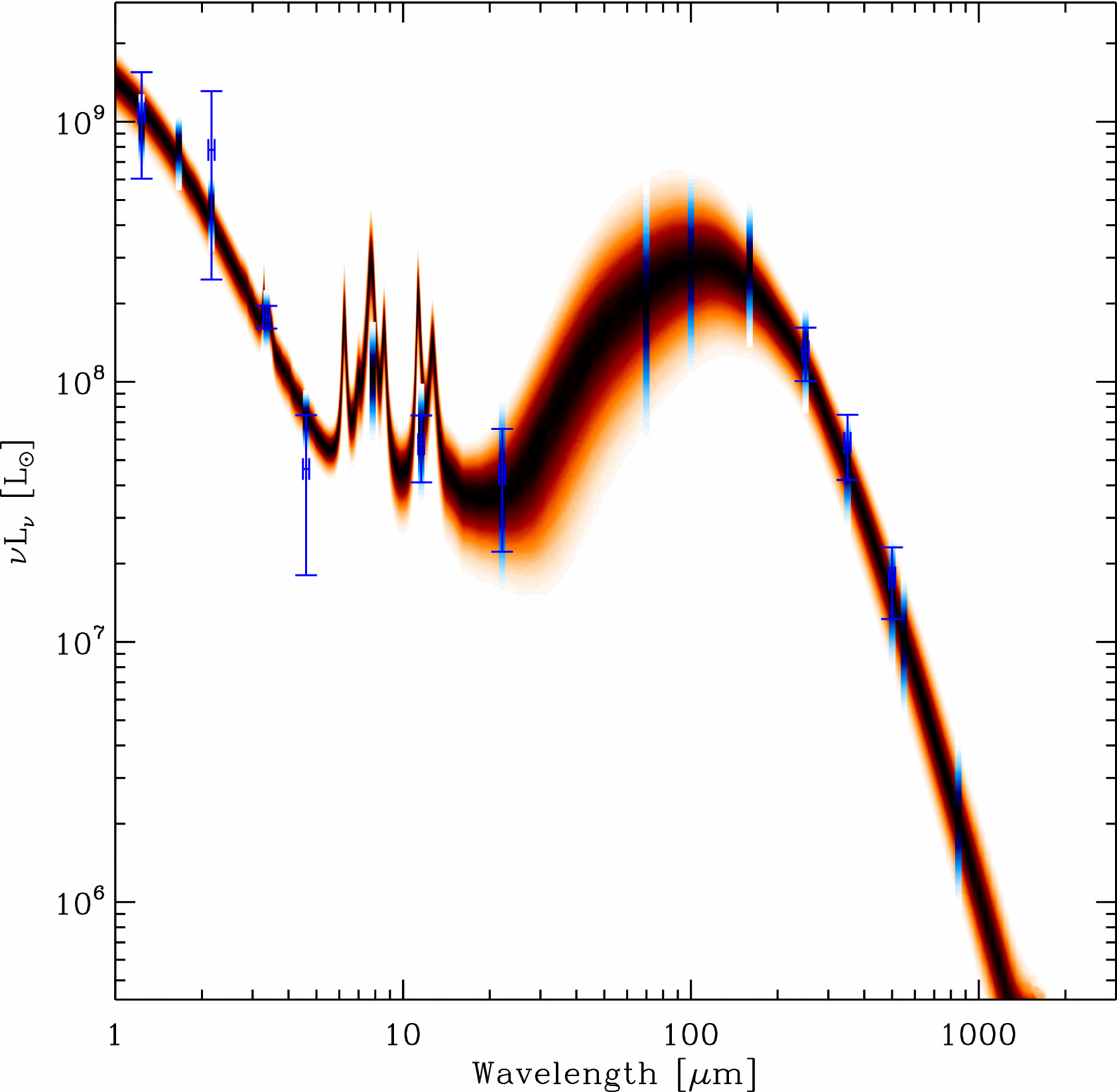}
      \caption{Dust-SEDs of selected individual galaxies in our sample. The observed data points are marked with the blue symbols, along with the uncertainty, while the model dust-SEDs are shown with the red, which represent the model dust-SEDs after each of the MCMC simulations performed. The blue-black shaded vertical symbols correspond to the synthetic photometry. The galaxies shown are NGC\,3485 (upper left), NGC4629 (upper right), UGC12160 (lower left), and UGC12709 (lower right).}
      \label{sl_figA3}%
 \end{figure*}
%
 We show the observed and model dust-SEDs for individual galaxies in Fig.~\ref{sl_figA3}. The galaxies for which the SEDs are shown are chosen to cover a range in wavelength coverage and signal-to-noise ratio, and correspond to NGC\,3485 (upper left panel), NGC4629 (upper right panel), UGC12160 (lower left panel), and UGC12709 (lower right panel). UGC12160 in the lower left panel is an example galaxy where observations in all bands (a total of 17; see Table~\ref{sl_table1}) exist. The shape of its model dust-SED in the FIR/submm part of the spectrum is constrained by the SPIRE\,350 \& 500$\mu$m upper limits, while the flux densities in the {\it Planck} bands are overestimated as compared to the SPIRE ones, possibly due to contamination by Galactic cirrus emission \citep{Planck2013XXVIII,Planck2016}.
 
The model dust-SEDs for all galaxies in our sample are listed in Table~\ref{sl_tableA2} (only the first 50 entries in Table~\ref{sl_tableA2} are shown here, while the full list of wavelengths (548 rows) and galaxies (753 columns) is available in the online version of the journal). Each row in Table~\ref{sl_tableA2} corresponds to one wavelength, $\lambda$ in $\mu$m, while each column (counting from the second one) corresponds to the $\nu$\,L$_{\nu}$\,(L$_{\sun}$) for one galaxy (see galaxy list in Table~\ref{sl_tableA1}). There are 548 rows (or wavelengths) and 754 columns (or 753 galaxies). The first column lists the wavelengths. Here, we list only the first 12 galaxies. 

\begin{sidewaystable*} 
\tiny
\caption[]{Galaxy sample and derived physical properties from the model and the calibrations.} 
\label{sl_tableA1} 
\centering    
\begin{tabular}{lrrrrrrrrrrrrrrr}
\hline\hline
Galaxy			&RA (J2000)		&Dec (J2000)		&T		&D25		&Distance		&M$_{dust}$		&$\sigma _{M_{dust}}$	&U$_{av}$	&$\sigma _{U_{av}}$ &QPAH	&$\sigma _{QPAH}$ &SFR$_{pass} ^{a}$ 	&$\sigma _{SFR}$ &log$_{10} (M_{\star}/M_{\sun}$) &$\sigma _{log _{10} (M_{\star}/M_{\sun} )}$         \\
			&(degrees)		&(degrees)		&		&arcmin	        &Mpc			&10$^{6}$\,M$_{\sun}$	&10$^{6}$\,M$_{\sun}$		&U$_{\sun}$	&U$_{\sun}$	  &	&	         &M$_{\sun}$/yr		&M$_{\sun}$/yr	&	                      &    	\\		
\hline
ESO097-013		&213.29130		&-65.33900		&3.3		&8.710		&4.20727		&3.54			&0.68			&25.69		&9.04		&0.074		&0.013		&6.062		&5.888		&10.568		&0.209			\\
ESO149-001		&356.95290		&-57.07261		&8.0		&3.388		&20.89297		&2.08			&1.03			&1.14		&0.86		&0.101		&0.035		&0.151		&0.135		&9.858		&0.209			\\	
ESO149-013		&0.69450		&-52.77158		&9.9		&1.175		&20.24850		&0.03			&0.04			&1.03		&1.13		&0.109		&0.037		&...		&...		&8.122		&0.207			\\
ESO157-047		&69.83010		&-54.21068		&0.0		&1.259		&21.49099		&0.05			&0.07			&1.14		&1.19		&0.114		&0.035		&3.088		&2.942		&8.457		&0.208			\\
ESO157-049		&69.90360		&-53.01267		&5.3		&1.738		&23.65920		&2.90			&1.10			&6.50		&2.70		&0.097		&0.011		&0.776		&0.714		&9.701		&0.209			\\
ESO209-009		&119.56440		&-49.85424		&6.1		&6.457		&12.58925		&13.39			&2.36			&2.84		&0.80		&0.132		&0.017		&1.365		&1.275		&10.312		&0.209			\\
ESO240-004		&351.97110		&-51.13189		&6.7		&1.778		&20.60631		&0.38			&0.32			&0.99		&0.92		&0.092		&0.031		&0.042		&0.036		&8.260		&0.207			\\
ESO351-030		&15.03885		&-33.70900		&-4.8		&1.148		&0.08433		&0.00			&0.00			&0.87		&1.57		&0.091		&0.059		&...		&...		&1.569		&0.201			\\
ESO358-015		&53.27805		&-34.80702		&8.7		&1.071		&16.14359		&0.03			&0.05			&1.09		&1.15		&0.112		&0.035		&1.829		&1.719		&8.219		&0.207			\\
ESO358-016		&53.28885		&-35.71908		&7.1		&1.122		&22.28291		&0.25			&0.28			&1.36		&1.37		&0.101		&0.029		&0.027		&0.023		&8.297		&0.207			\\
ESO358-019		&53.62860		&-34.29750		&-3.0		&1.175		&16.04315		&0.07			&0.11			&1.15		&1.25		&0.109		&0.033		&...		&...		&8.232		&0.207			\\
ESO358-042		&54.53820		&-34.51858		&-2.2		&1.230		&20.41738		&0.02			&0.02			&1.16		&1.29		&0.125		&0.038		&...		&...		&8.470		&0.208			\\
ESO358-043		&54.55680		&-33.12725		&-3.0		&1.318		&20.79696		&0.60			&0.43			&0.65		&0.54		&0.114		&0.030		&0.018		&0.015		&8.994		&0.208			\\
ESO358-050		&55.26510		&-33.77946		&-2.1		&1.622		&19.67885		&0.18			&0.17			&1.42		&1.35		&0.131		&0.033		&0.013		&0.012		&9.702		&0.209			\\
ESO358-051		&55.38585		&-34.88838		&0.3		&1.413		&14.30000		&0.81			&0.53			&2.08		&1.41		&0.093		&0.014		&0.075		&0.065		&8.837		&0.208			\\
ESO358-054		&55.75965		&-36.27269		&7.9		&1.660		&16.90000		&1.61			&0.74			&0.43		&0.27		&0.096		&0.026		&0.042		&0.036		&8.714		&0.208			\\
ESO358-056		&55.84425		&-33.94002		&-2.2		&1.413		&21.37962		&0.04			&0.05			&1.43		&1.51		&0.120		&0.034		&...		&...		&8.670		&0.208			\\
ESO358-059		&56.26500		&-35.97270		&-3.3		&1.380		&19.58844		&0.04			&0.05			&1.67		&1.82		&0.137		&0.035		&2.605		&2.472		&9.379		&0.208			\\
ESO358-060		&56.30130		&-35.57055		&9.8		&1.622		&15.70000		&0.01			&0.03			&0.94		&1.09		&0.099		&0.039		&...		&...		&7.456		&0.207			\\
ESO358-063		&56.57910		&-34.94359		&6.9		&4.266		&16.82673		&5.96			&0.52			&3.48		&0.41		&0.129		&0.009		&0.768		&0.708		&9.989		&0.209			\\
ESO358-066		&56.96940		&-36.47167		&-3.8		&1.097		&19.76970		&0.05			&0.07			&1.38		&1.40		&0.115		&0.033		&0.005		&0.004		&8.527		&0.208			\\
ESO359-002		&57.65295		&-35.90937		&-3.4		&1.259		&19.14256		&0.10			&0.13			&2.23		&2.39		&0.116		&0.033		&...		&...		&9.183		&0.208			\\
ESO373-008		&143.33955		&-33.03346		&6.1		&3.715		&9.68278		&5.08			&1.15			&1.17		&0.40		&0.102		&0.018		&0.241		&0.216		&9.498		&0.209			\\
ESO377-039		&169.84035		&-36.59571		&3.3		&1.071		&36.29164		&1.08			&0.76			&1.38		&1.10		&0.116		&0.027		&0.076		&0.066		&9.463		&0.208			\\
ESO405-014		&335.69580		&-34.83391		&6.1		&1.288		&36.83779		&0.10			&0.16			&1.34		&1.48		&0.097		&0.035		&...		&...		&...		&...  			\\
ESO406-031		&344.41995		&-35.39701		&2.2		&1.479		&22.52867		&0.72			&0.34			&1.83		&1.05		&0.119		&0.023		&0.048		&0.042		&9.228		&0.208			\\
ESO407-002		&346.17270		&-34.05781		&-0.5		&1.097		&23.96232		&1.12			&0.26			&5.07		&1.58		&0.120		&0.015		&0.217		&0.193		&9.559		&0.209			\\
ESO407-018		&351.61605		&-32.38858		&9.8		&1.349		&2.22844		&0.00			&0.00			&0.96		&1.14		&0.114		&0.042		&...		&...		&7.311		&0.207			\\
ESO410-012		&7.07520		&-27.98335		&4.7		&1.413		&21.01311		&0.04			&0.06			&1.22		&1.42		&0.098		&0.037		&0.017		&0.014		&7.949		&0.207			\\
ESO410-018		&8.54625		&-30.77392		&8.9		&1.549		&19.00000		&1.22			&0.61			&1.04		&0.67		&0.104		&0.024		&0.061		&0.073		&8.810		&0.208			\\
ESO411-013		&11.77665		&-31.58135		&9.0		&1.230		&23.78482		&0.29			&0.26			&1.49		&1.43		&0.122		&0.033		&...		&...		&...		&...  			\\
ESO411-016		&11.91495		&-27.94758		&1.8		&1.023		&24.08520		&0.46			&0.45			&0.67		&0.61		&0.112		&0.030		&...		&...		&8.632		&0.208			\\
ESO411-026		&13.18965		&-31.71818		&9.0		&1.148		&21.25887		&0.05			&0.09			&0.90		&0.96		&0.106		&0.036		&...		&...		&8.188		&0.207			\\
ESO411-027		&13.21500		&-27.32558		&5.7		&1.259		&24.65866		&0.15			&0.20			&1.14		&1.17		&0.097		&0.033		&...		&...		&8.080		&0.207			\\
ESO428-014		&109.12995		&-29.32481		&-1.8		&1.995		&23.20000		&4.73			&3.17			&16.68		&11.82		&0.083		&0.019		&5.645		&5.470		&10.612		&0.210			\\
ESO434-040		&146.91765		&-30.94863		&-1.3		&1.071		&35.48607		&0.00			&0.00			&76110.00	&17640.00	&0.071		&0.002		&14.012		&13.888		&10.797		&0.210			\\
ESO493-016		&117.18255		&-26.24659		&4.0		&1.000		&23.97000		&15.71			&1.92			&7.68		&0.93		&0.094		&0.006		&4.633		&4.467		&10.422		&0.209			\\
ESO495-021		&129.06330		&-26.40933		&-2.6		&1.995		&8.24138		&0.41			&0.01			&53.89		&1.79		&0.061		&0.004		&2.627		&2.492		&9.367		&0.208			\\
ESO549-035		&58.76625		&-20.38338		&6.0		&1.023		&38.90453		&1.58			&0.98			&0.81		&0.63		&0.098		&0.031		&9.085		&8.965		&8.964		&0.208			\\
ESO560-004		&114.75195		&-22.04626		&4.2		&1.380		&37.49729		&9.72			&1.64			&8.27		&1.60		&0.097		&0.009		&2.509		&2.390		&10.700		&0.210			\\
ESO603-031		&344.36010		&-17.88565		&1.0		&1.122		&35.48135		&1.72			&0.87			&1.74		&1.13		&0.109		&0.024		&0.143		&0.126		&9.393		&0.208			\\
IC0010			&5.09625		&59.29310		&9.9		&6.607		&0.79433		&0.21			&0.07			&4.93		&1.59		&0.016		&0.012		&0.066		&0.057		&8.726		&0.208			\\
IC0335			&53.87895		&-34.44708		&-2.1		&2.630		&20.51161		&0.16			&0.14			&2.88		&2.71		&0.142		&0.031		&0.019		&0.020		&9.953		&0.209			\\
IC0342			&56.70540		&68.10139		&6.0		&19.953		&3.43558		&4.78			&0.32			&15.43		&1.05		&0.139		&0.005		&3.256		&3.113		&10.532		&0.209			\\
IC0610			&156.61830		&20.22826		&3.7		&1.905		&29.24156		&4.83			&0.54			&5.40		&0.93		&0.088		&0.008		&0.746		&0.687		&10.102		&0.209			\\
\hline
\end{tabular} 							
\tablefoot{This table is available in its entirety in the online journal, while here the first 45 entries are listed to demonstrate its form and content.\\							
\tablefoottext{a}{The SFR have been derived from flux densities corrected for the evolved stars contribution (see Section~2.2).}\\
}
\end{sidewaystable*}

\begin{sidewaystable*}
\tiny
\caption[]{Dust-SEDs for the galaxy sample. The first column corresponds to the wavelength $\lambda$\,($\mu$m), while subsequent columns correspond to the $\nu$\,L$_{\nu}$ (L$_{\sun}$) for each galaxy in our sample.}
\label{sl_tableA2}
\centering   
\begin{tabular}{rrrrrrrrrrrrr}
\hline\hline
$\lambda$\,($\mu$m)	&ESO097-013	&ESO149-001	&ESO149-013	&ESO157-047	&ESO157-049	&ESO209-009	&ESO240-004	&ESO351-030	&ESO358-015	&ESO358-016	&ESO358-019	&ESO358-042     \\
\hline
1.006        &+3.087e+10     &+5.933e+09     &+1.817e+08     &+3.785e+08     &+3.472e+09     &+1.120e+10     &+2.250e+08     &+2.933e+02     &+1.893e+08     &+2.622e+08     &+2.172e+08     &+4.556e+08     \\
1.012        &+3.072e+10     &+5.901e+09     &+1.807e+08     &+3.765e+08     &+3.451e+09     &+1.114e+10     &+2.237e+08     &+2.916e+02     &+1.882e+08     &+2.607e+08     &+2.160e+08     &+4.529e+08     \\
1.019	     &+3.051e+10     &+5.861e+09     &+1.795e+08     &+3.739e+08     &+3.430e+09     &+1.107e+10     &+2.223e+08     &+2.896e+02     &+1.869e+08     &+2.590e+08     &+2.145e+08     &+4.498e+08     \\
1.025	     &+3.033e+10     &+5.826e+09     &+1.784e+08     &+3.717e+08     &+3.410e+09     &+1.100e+10     &+2.210e+08     &+2.880e+02     &+1.858e+08     &+2.575e+08     &+2.133e+08     &+4.472e+08     \\
1.028	     &+3.024e+10     &+5.809e+09     &+1.779e+08     &+3.707e+08     &+3.400e+09     &+1.097e+10     &+2.203e+08     &+2.871e+02     &+1.853e+08     &+2.567e+08     &+2.126e+08     &+4.459e+08     \\
1.032	     &+3.012e+10     &+5.787e+09     &+1.772e+08     &+3.692e+08     &+3.387e+09     &+1.093e+10     &+2.195e+08     &+2.860e+02     &+1.846e+08     &+2.557e+08     &+2.118e+08     &+4.442e+08     \\
1.035	     &+3.004e+10     &+5.770e+09     &+1.767e+08     &+3.682e+08     &+3.377e+09     &+1.090e+10     &+2.188e+08     &+2.851e+02     &+1.840e+08     &+2.549e+08     &+2.112e+08     &+4.429e+08     \\
1.038	     &+2.995e+10     &+5.750e+09     &+1.761e+08     &+3.671e+08     &+3.368e+09     &+1.087e+10     &+2.182e+08     &+2.843e+02     &+1.835e+08     &+2.542e+08     &+2.105e+08     &+4.416e+08     \\
1.041	     &+2.986e+10     &+5.734e+09     &+1.756e+08     &+3.660e+08     &+3.358e+09     &+1.083e+10     &+2.175e+08     &+2.834e+02     &+1.829e+08     &+2.534e+08     &+2.099e+08     &+4.403e+08     \\
1.045	     &+2.972e+10     &+5.712e+09     &+1.749e+08     &+3.643e+08     &+3.342e+09     &+1.079e+10     &+2.167e+08     &+2.823e+02     &+1.822e+08     &+2.524e+08     &+2.091e+08     &+4.386e+08     \\
1.046	     &+2.969e+10     &+5.706e+09     &+1.747e+08     &+3.640e+08     &+3.339e+09     &+1.078e+10     &+2.164e+08     &+2.820e+02     &+1.820e+08     &+2.521e+08     &+2.088e+08     &+4.379e+08     \\
1.048	     &+2.964e+10     &+5.693e+09     &+1.744e+08     &+3.633e+08     &+3.333e+09     &+1.076e+10     &+2.160e+08     &+2.814e+02     &+1.816e+08     &+2.516e+08     &+2.084e+08     &+4.371e+08     \\
1.051	     &+2.955e+10     &+5.676e+09     &+1.739e+08     &+3.623e+08     &+3.323e+09     &+1.073e+10     &+2.153e+08     &+2.806e+02     &+1.811e+08     &+2.509e+08     &+2.078e+08     &+4.359e+08     \\
1.053	     &+2.950e+10     &+5.666e+09     &+1.735e+08     &+3.616e+08     &+3.317e+09     &+1.070e+10     &+2.149e+08     &+2.800e+02     &+1.807e+08     &+2.503e+08     &+2.073e+08     &+4.350e+08     \\
1.054	     &+2.947e+10     &+5.660e+09     &+1.733e+08     &+3.609e+08     &+3.314e+09     &+1.069e+10     &+2.147e+08     &+2.797e+02     &+1.805e+08     &+2.501e+08     &+2.071e+08     &+4.346e+08     \\
1.058	     &+2.936e+10     &+5.636e+09     &+1.726e+08     &+3.596e+08     &+3.301e+09     &+1.065e+10     &+2.138e+08     &+2.786e+02     &+1.798e+08     &+2.491e+08     &+2.063e+08     &+4.327e+08     \\
1.059	     &+2.930e+10     &+5.631e+09     &+1.724e+08     &+3.592e+08     &+3.295e+09     &+1.064e+10     &+2.136e+08     &+2.783e+02     &+1.796e+08     &+2.488e+08     &+2.061e+08     &+4.323e+08     \\
1.061	     &+2.924e+10     &+5.617e+09     &+1.721e+08     &+3.586e+08     &+3.289e+09     &+1.062e+10     &+2.131e+08     &+2.777e+02     &+1.792e+08     &+2.483e+08     &+2.057e+08     &+4.315e+08     \\
1.064	     &+2.916e+10     &+5.601e+09     &+1.716e+08     &+3.576e+08     &+3.280e+09     &+1.059e+10     &+2.125e+08     &+2.769e+02     &+1.787e+08     &+2.476e+08     &+2.051e+08     &+4.302e+08     \\
1.068	     &+2.905e+10     &+5.580e+09     &+1.709e+08     &+3.559e+08     &+3.267e+09     &+1.054e+10     &+2.116e+08     &+2.758e+02     &+1.780e+08     &+2.466e+08     &+2.042e+08     &+4.284e+08     \\
1.071	     &+2.894e+10     &+5.562e+09     &+1.703e+08     &+3.549e+08     &+3.255e+09     &+1.051e+10     &+2.110e+08     &+2.749e+02     &+1.774e+08     &+2.458e+08     &+2.036e+08     &+4.272e+08     \\
1.078	     &+2.876e+10     &+5.523e+09     &+1.692e+08     &+3.524e+08     &+3.234e+09     &+1.044e+10     &+2.095e+08     &+2.730e+02     &+1.762e+08     &+2.441e+08     &+2.022e+08     &+4.241e+08     \\
1.085	     &+2.854e+10     &+5.485e+09     &+1.679e+08     &+3.498e+08     &+3.211e+09     &+1.036e+10     &+2.080e+08     &+2.711e+02     &+1.749e+08     &+2.423e+08     &+2.007e+08     &+4.211e+08     \\
1.091	     &+2.836e+10     &+5.449e+09     &+1.669e+08     &+3.476e+08     &+3.190e+09     &+1.030e+10     &+2.067e+08     &+2.693e+02     &+1.738e+08     &+2.408e+08     &+1.994e+08     &+4.182e+08     \\
1.098	     &+2.815e+10     &+5.409e+09     &+1.657e+08     &+3.451e+08     &+3.167e+09     &+1.023e+10     &+2.052e+08     &+2.674e+02     &+1.725e+08     &+2.391e+08     &+1.980e+08     &+4.153e+08     \\
1.105	     &+2.794e+10     &+5.369e+09     &+1.644e+08     &+3.427e+08     &+3.144e+09     &+1.015e+10     &+2.037e+08     &+2.654e+02     &+1.712e+08     &+2.373e+08     &+1.965e+08     &+4.121e+08     \\
1.109	     &+2.782e+10     &+5.344e+09     &+1.637e+08     &+3.409e+08     &+3.130e+09     &+1.010e+10     &+2.027e+08     &+2.641e+02     &+1.704e+08     &+2.362e+08     &+1.956e+08     &+4.104e+08     \\
1.112	     &+2.771e+10     &+5.327e+09     &+1.631e+08     &+3.400e+08     &+3.119e+09     &+1.007e+10     &+2.021e+08     &+2.632e+02     &+1.699e+08     &+2.354e+08     &+1.949e+08     &+4.090e+08     \\
1.119	     &+2.754e+10     &+5.289e+09     &+1.620e+08     &+3.376e+08     &+3.097e+09     &+1.000e+10     &+2.007e+08     &+2.614e+02     &+1.687e+08     &+2.338e+08     &+1.936e+08     &+4.062e+08     \\
1.126	     &+2.732e+10     &+5.250e+09     &+1.608e+08     &+3.349e+08     &+3.075e+09     &+9.926e+09     &+1.992e+08     &+2.595e+02     &+1.674e+08     &+2.320e+08     &+1.921e+08     &+4.031e+08     \\
1.133	     &+2.712e+10     &+5.210e+09     &+1.596e+08     &+3.323e+08     &+3.051e+09     &+9.854e+09     &+1.977e+08     &+2.575e+02     &+1.662e+08     &+2.303e+08     &+1.907e+08     &+4.001e+08     \\
1.147	     &+2.674e+10     &+5.139e+09     &+1.574e+08     &+3.278e+08     &+3.011e+09     &+9.720e+09     &+1.950e+08     &+2.540e+02     &+1.639e+08     &+2.271e+08     &+1.881e+08     &+3.944e+08     \\
1.162	     &+2.632e+10     &+5.057e+09     &+1.548e+08     &+3.225e+08     &+2.962e+09     &+9.564e+09     &+1.918e+08     &+2.499e+02     &+1.613e+08     &+2.235e+08     &+1.851e+08     &+3.883e+08     \\
1.176	     &+2.590e+10     &+4.979e+09     &+1.525e+08     &+3.176e+08     &+2.916e+09     &+9.419e+09     &+1.889e+08     &+2.461e+02     &+1.588e+08     &+2.201e+08     &+1.822e+08     &+3.821e+08     \\
1.191	     &+2.550e+10     &+4.896e+09     &+1.500e+08     &+3.124e+08     &+2.870e+09     &+9.266e+09     &+1.858e+08     &+2.420e+02     &+1.562e+08     &+2.164e+08     &+1.792e+08     &+3.758e+08     \\
1.206	     &+2.506e+10     &+4.815e+09     &+1.475e+08     &+3.072e+08     &+2.824e+09     &+9.116e+09     &+1.827e+08     &+2.380e+02     &+1.536e+08     &+2.129e+08     &+1.763e+08     &+3.696e+08     \\
1.221	     &+2.465e+10     &+4.736e+09     &+1.450e+08     &+3.020e+08     &+2.777e+09     &+8.964e+09     &+1.797e+08     &+2.340e+02     &+1.510e+08     &+2.093e+08     &+1.733e+08     &+3.634e+08     \\
1.252	     &+2.392e+10     &+4.595e+09     &+1.407e+08     &+2.933e+08     &+2.696e+09     &+8.704e+09     &+1.744e+08     &+2.271e+02     &+1.466e+08     &+2.032e+08     &+1.682e+08     &+3.527e+08     \\
1.284	     &+2.308e+10     &+4.434e+09     &+1.358e+08     &+2.830e+08     &+2.601e+09     &+8.401e+09     &+1.683e+08     &+2.192e+02     &+1.414e+08     &+1.961e+08     &+1.623e+08     &+3.404e+08     \\
1.317	     &+2.224e+10     &+4.273e+09     &+1.308e+08     &+2.727e+08     &+2.506e+09     &+8.099e+09     &+1.622e+08     &+2.112e+02     &+1.363e+08     &+1.889e+08     &+1.564e+08     &+3.280e+08     \\
1.350	     &+2.142e+10     &+4.115e+09     &+1.260e+08     &+2.625e+08     &+2.416e+09     &+7.803e+09     &+1.562e+08     &+2.034e+02     &+1.312e+08     &+1.820e+08     &+1.506e+08     &+3.158e+08     \\
1.384	     &+2.060e+10     &+3.958e+09     &+1.212e+08     &+2.526e+08     &+2.324e+09     &+7.510e+09     &+1.503e+08     &+1.956e+02     &+1.262e+08     &+1.750e+08     &+1.449e+08     &+3.037e+08     \\
1.419	     &+1.979e+10     &+3.803e+09     &+1.164e+08     &+2.425e+08     &+2.233e+09     &+7.219e+09     &+1.444e+08     &+1.879e+02     &+1.213e+08     &+1.682e+08     &+1.392e+08     &+2.918e+08     \\
1.455	     &+1.899e+10     &+3.649e+09     &+1.117e+08     &+2.328e+08     &+2.145e+09     &+6.931e+09     &+1.386e+08     &+1.804e+02     &+1.164e+08     &+1.614e+08     &+1.336e+08     &+2.800e+08     \\
1.492	     &+1.820e+10     &+3.498e+09     &+1.071e+08     &+2.232e+08     &+2.058e+09     &+6.649e+09     &+1.329e+08     &+1.729e+02     &+1.116e+08     &+1.548e+08     &+1.281e+08     &+2.684e+08     \\
1.530	     &+1.743e+10     &+3.351e+09     &+1.026e+08     &+2.138e+08     &+1.971e+09     &+6.370e+09     &+1.273e+08     &+1.656e+02     &+1.068e+08     &+1.482e+08     &+1.226e+08     &+2.571e+08     \\
1.569	     &+1.668e+10     &+3.206e+09     &+9.813e+07     &+2.044e+08     &+1.886e+09     &+6.099e+09     &+1.218e+08     &+1.584e+02     &+1.022e+08     &+1.418e+08     &+1.173e+08     &+2.459e+08     \\
1.609	     &+1.594e+10     &+3.063e+09     &+9.379e+07     &+1.955e+08     &+1.804e+09     &+5.834e+09     &+1.164e+08     &+1.514e+02     &+9.771e+07     &+1.356e+08     &+1.121e+08     &+2.351e+08     \\
1.650	     &+1.523e+10     &+2.925e+09     &+8.957e+07     &+1.866e+08     &+1.724e+09     &+5.574e+09     &+1.112e+08     &+1.446e+02     &+9.330e+07     &+1.295e+08     &+1.071e+08     &+2.246e+08     \\
1.691	     &+1.454e+10     &+2.794e+09     &+8.552e+07     &+1.782e+08     &+1.646e+09     &+5.326e+09     &+1.062e+08     &+1.381e+02     &+8.909e+07     &+1.237e+08     &+1.023e+08     &+2.143e+08     \\
\hline
\end{tabular}
\tablefoot{This table is available in its entirety in the online journal. The first 50 entries are shown here for guidance regarding its form and content.}
\end{sidewaystable*}
%

\section{Dust-SED properties using modified black bodies}

\begin{figure*}
 \centering
 \includegraphics[width=7.7cm,clip]{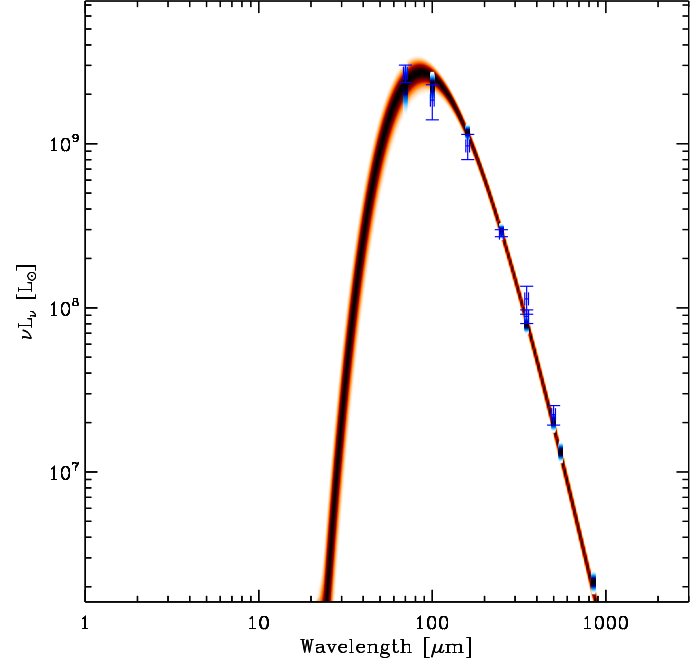}
 \includegraphics[width=7.7cm,clip]{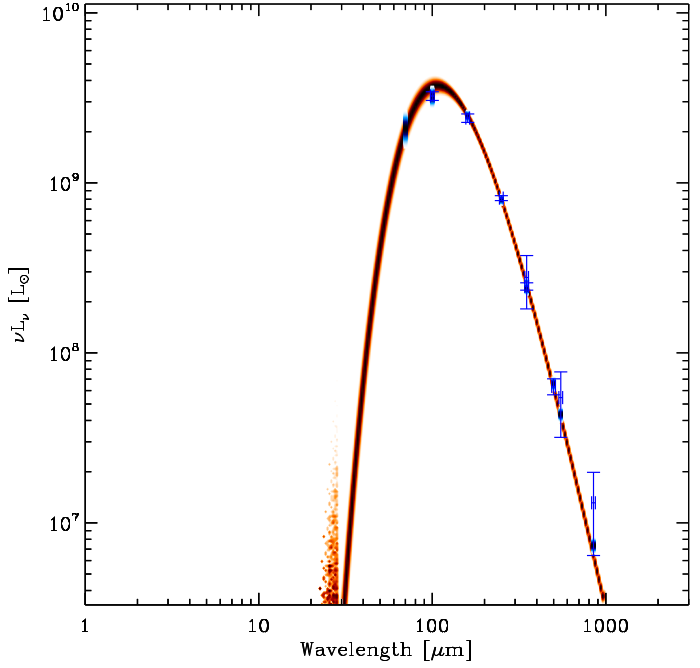}
 \caption{Modified black body SEDs of individual galaxies. The observed data points are marked with the blue symbols, along with the uncertainty, while the model dust-SEDs are shown with the red, representing the dust-SEDs after each of the MCMC simulations performed. The blue-black shaded vertical symbols correspond to the synthetic photometry. The galaxies for which the dust SEDs are shown correspond to NGC\,3353 (left panel) and NGC\,4607 (right panel).}
      \label{sl_figB1}%
 \end{figure*}
%
In addition to the non-uniformly illuminated dust component, we model the galaxy sample with simple modified black bodies, within the hierarchical Bayesian framework. Example modified black body model dust-SEDs for individual galaxies, with the observed data points overplotted, are shown in Fig.~\ref{sl_figB1}. The SEDs correspond to the galaxies NGC\,3353 (left panel) and NGC\,4607 (right panel).

\begin{table}
\tiny
      \caption[]{Dust masses and temperatures of the modified blackbody dust-SED modelling for the same galaxy sample listed in Table A.1. } 
      \label{sl_tableB1}
      \centering   
      \begin{tabular}{rrrrr}
\hline\hline
Galaxy			&M$_{dust}$		&$\sigma _{M_{dust}}$	&T	&$\sigma _{T}$  \\
			&10$^{6}\times$M$_{\sun}$  &10$^{6}\times$M$_{\sun}$ &K	&K	        \\		
\hline
ESO097-013		&5.28		&1.16		&26		&3  	\\
ESO149-001		&1.69		&0.67		&20		&3  	\\	
ESO149-013		&0.06		&0.09		&18		&4  	\\
ESO157-047		&0.18		&0.28		&18		&3  	\\
ESO157-049		&2.76		&0.98		&25		&2  	\\
ESO209-009		&14.93		&3.04		&20		&2  	\\
ESO240-004		&0.33		&0.22		&19		&3  	\\
ESO351-030		&0.00		&0.00		&9		&4  	\\
ESO358-015		&0.04		&0.05		&17		&4  	\\
ESO358-016		&0.13		&0.21		&18		&3  	\\
\hline
\end{tabular}
\tablefoot{This table is available in its entirety in the online journal. The first 10 entries are shown here for guidance regarding its form and content.}
\end{table}
%
The derived properties of the modified black body modelling of the full sample of the galaxies, which were modelled as an ensemble, is shown in Table~\ref{sl_tableB1} (the full list of the galaxies is available in the online version of the journal). The columns show:
   (1) the galaxy name;
   (2) the model-derived dust mass, M$_{dust}$ in M$_{\sun}$, for the THEMIS dust grain model;
   (3) the standard deviation of the dust mass, $\sigma _{M_{dust}}$ in M$_{\sun}$;
   (4) the dust temperature, T in K; 
   (5) the standard deviation of the dust temperature, $\sigma _{T}$ in K.

\begin{table}
\tiny
\caption[]{Modified black body SEDs for the galaxy sample listed in Table A.1. The first column corresponds to the wavelength, $\lambda$\,($\mu$m), while subsequent columns correspond to the $\nu$\,L$_{\nu}$ (L$_{\sun}$) for individual galaxies. } 
\label{sl_tableB2}
\centering   
\begin{tabular}{rrrrr}
\hline\hline
$\lambda$\,($\mu$m)	&ESO097-013  &ESO149-001  &ESO149-013	      &ESO157-047    \\
\hline
1.006	&+3.871e-214     &0               &0               &0                  \\
1.012	&+1.129e-212     &0               &0               &0                  \\
1.019	&+3.222e-211     &0               &0               &0                  \\
1.025	&+9.010e-210     &0               &0               &0                  \\
1.028	&+4.726e-209     &0               &0               &0                  \\
1.032	&+2.466e-208     &0               &0               &0                  \\
1.035	&+1.280e-207     &0               &0               &0                  \\
1.038	&+6.608e-207     &+1.507e-284     &0               &0                  \\
1.041	&+3.394e-206     &+1.349e-283     &0               &0                  \\
1.045	&+1.735e-205     &+1.198e-282     &0               &0                  \\
1.046	&+3.914e-205     &+3.562e-282     &0               &0                  \\
1.048	&+8.821e-205     &+1.057e-281     &0               &0                  \\
1.051	&+4.464e-204     &+9.266e-281     &0               &0                  \\
1.053	&+1.002e-203     &+2.736e-280     &0               &0                  \\
1.054	&+2.246e-203     &+8.066e-280     &0               &0                  \\
1.058	&+1.124e-202     &+6.972e-279     &0               &0                  \\
1.059	&+2.511e-202     &+2.045e-278     &0               &0                  \\
1.061	&+5.600e-202     &+5.987e-278     &0               &0                  \\
1.064	&+2.776e-201     &+5.107e-277     &0               &0                  \\
1.068	&+1.369e-200     &+4.327e-276     &0               &0                  \\
\hline
\end{tabular}
\tablefoot{This table is available in its entirety in the online journal. The first 40 rows and 5 columns are shown here for guidance on its content.\\
A very small or zero entry in the columns for the $\nu$\,L$_{\nu}$ of each galaxy occurs in the short wavelength regime, for the portion of the table shown here.}
\end{table}
%
The modified black body dust-SEDs for all galaxies studied here are shown in Table~\ref{sl_tableB2} (the full list of wavelengths (548 rows) and galaxies (753 columns) is available in the online version of the journal). Each row in Table~\ref{sl_tableB2} corresponds to one wavelength, $\lambda$ in $\mu$m, while each column (counting from the second one) corresponds to $\nu$\,L$_{\nu}$ (L$_{\sun}$) for one galaxy (see Table~\ref{sl_tableA1} for the galaxy sample). There are 548 rows and 754 columns. The first column lists the wavelength. Here, we list only the first 4 galaxies.

 \begin{figure*}[th!]
  \centering
      \includegraphics[width=7.5cm,clip]{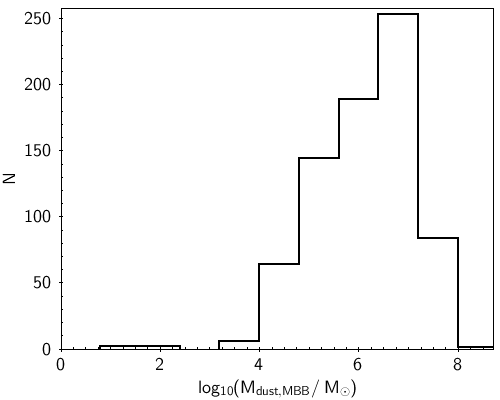}
      \includegraphics[width=7.5cm,clip]{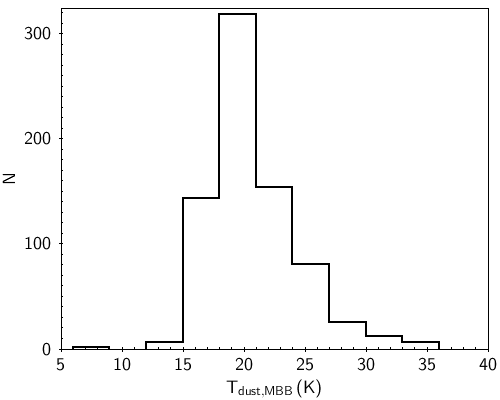}
      \includegraphics[width=7.5cm,clip]{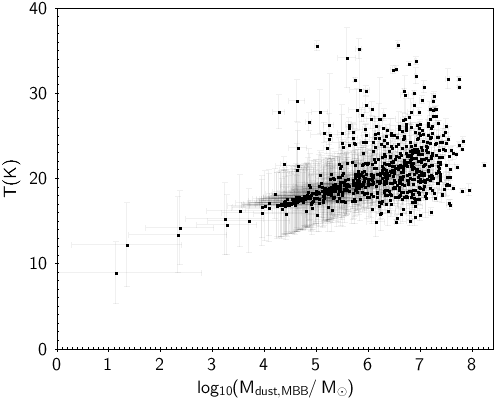}
     \includegraphics[width=7.5cm,clip]{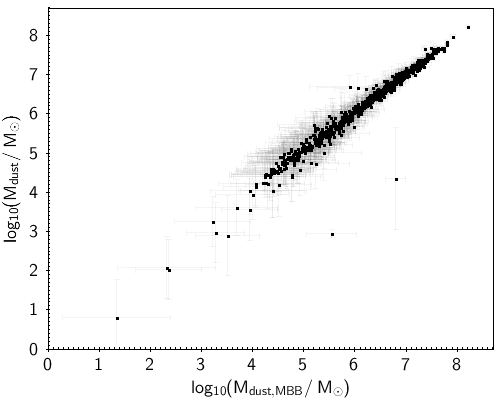}
      \caption{{\it Upper panels}: Distribution of dust masses (left) and temperatures (right) from fitting modified black bodies to the galaxy sample. 
{\it Lower panels}: Temperature versus dust masses from fitting modified black bodies (left). Dust masses from the non-uniformly illuminated dust mixture versus the modified black body dust masses (right).
      } 
      \label{sl_figB2}%
 \end{figure*}
%
 Fig.~\ref{sl_figB2} shows the distribution of the modified black body dust masses (upper left) and  temperatures (upper right). The relation between them is shown in the lower left panel of the same figure. A comparison between the dust masses from the non-uniformly illuminated dust mixture and the modified black body dust masses is shown in the lower right panel of Fig.~\ref{sl_figB2}. We note that there is remarkable consistency between the two model choices of fitting the dust-SEDs. Temperatures are not derived in the dust-SED model case of the non-uniformly illuminated dust component.    

\section{Dust specific SFR versus U$_{av}$}    
        
 \begin{figure}
  \centering
      \includegraphics[width=8cm,clip]{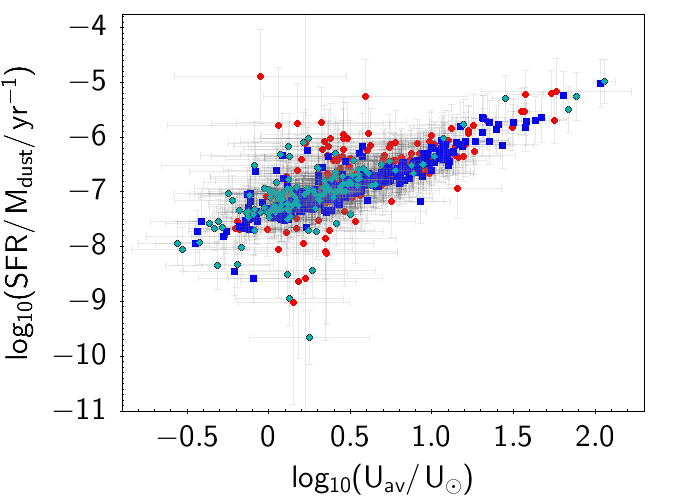}
      \caption{Ratio of SFR over M$_{dust}$ as a function of U$_{av}$. The error bars represent the standard deviation for the U$_{av}$, and the propagated uncertainty for the ratio of SFR over M$_{dust}$.}
      \label{sl_figC1}
 \end{figure}
 Fig.~\ref{sl_figC1} shows the relation between the ratio of the SFR-to-M$_{dust}$ (i.e., the dust-specific SFR) with U$_{av}$. M$_{dust}$ traces the gas mass, via a GDR. Hence, the ratio of the SFR over the dust mass is a tracer of the SFE. This SFE-proxy correlates with the U$_{av}$, for more than two orders of magnitude in U$_{av}$, primarily holding for LTGs and Irs. The few ETGs that scatter to larger SFE-proxy values are the same as those discussed earlier (Section 6.1). Higher ISRF intensities, U$_{av}$ indicate harder photons coming from regions of ongoing massive star formation (i.e., higher SFE) and with an SED peaking to hotter dust temperatures.
 
\end{appendix}

\end{document}